\documentclass[11pt]{article}
\usepackage[latin1]{inputenc}
\usepackage[a4paper,scale={.83,.85}]{geometry}
\usepackage{amsmath,amssymb,amsfonts,graphicx,caption,float,multirow}
\usepackage[usenames, dvipsnames]{xcolor}
\usepackage[colorlinks=true, pdfstartview=FitV,linkcolor=ForestGreen,citecolor=ForestGreen, urlcolor=blue]{hyperref}
\usepackage[hang]{footmisc} 

\newcommand{\N}{\mathbb{N}}
 
\newcommand{\R}{\mathbb{R}}
\newcommand{\C}{\mathbb{C}} 
\newcommand{\T}{\mathbb{T}} 
\newcommand{\power}{\mathfrak{P}}
\newcommand{\eg}{\textit{e.g.~}}
\newcommand{\ie}{\textit{i.e.~}}

\newtheorem{thm}{Theorem}
\newtheorem{cor}[thm]{Corollary}
\newtheorem{prop}[thm]{Proposition}
\newtheorem{lemma}[thm]{Lemma}
\newtheorem{dfn}[thm]{Definition}
\newcommand{\Proof}{\paragraph{\it Proof.}}
\newcommand{\cqfd}{\hfill\rule{1ex}{1ex}}

\begin{document}
\title{On the Wiener-Khinchin transform of functions\\
that behave as approximate power-laws.\\
Applications to fluid turbulence.}
\author{François Vigneron\\[1ex]
\small\texttt{francois.vigneron@u-pec.fr}\\[.5ex]
\small Université Paris-Est\\ 
\small LAMA (\textsc{umr8050}), UPEC, UPEM, CNRS \\
\small 61, avenue du Général de Gaulle, F94010 Créteil, France.}
\maketitle

\begin{abstract}
As we all know, the Fourier transform is continuous in the weak sense of tempered distribution; this ensures the weak stability of Fourier pairs.
This article investigates a stronger form of stability of the pair of homogeneous profiles $(|x|^{-\alpha},c_d|\xi|^{d-\alpha})$ on $\mathbb{R}^d$.
It encompasses, for example, the case where the homogeneous profiles exist only on a large but finite range.
In this case, we provide precise error estimates in terms of the size of the tails outside the homogeneous range.
We also prove a series of refined properties of the Fourier transform on related questions including criteria that ensure an approximate
homogeneous behavior asymptotically near the origin or at infinity. The sharpness of our results is checked with numerical simulations.
We also investigate how these results consolidate the mathematical foundations of turbulence theory.\\[1ex]
\textbf{Keywords:} Wiener-Khinchin transform, Hankel transform, Quasi-power-law, Turbulence, Energy spectrum, Structure functions.\\[1ex]
\textbf{MSC primary:} 42B10\\
\textbf{MSC secondary:}  44A15, 65R10, 33C10, 76F02, 76F05.
\end{abstract}

\section{Introduction}

We are interested in the following integral transformation of $f\in L^1(0,+\infty)$:
\begin{equation}\label{main}
\mathbb{WK}_d[f](\lambda)=\int_0^\infty H_d(\lambda k)f(k)dk = \lambda^{-1} \int_0^\infty H_d(\sigma)f(\sigma/\lambda) d\sigma
\end{equation}
where the kernel $H_d$ is defined for $d\in\N^\ast$ by
\begin{equation}\label{WKkernel}
H_d(\sigma)=1-  \Gamma\left(\tfrac{d}{2}\right) \cdot (\pi\sigma)^{1-\frac{d}{2}} J_{\frac{d}{2}-1}(2\pi\sigma).
\end{equation}

This particular integral transform occurs in hydrodynamics : if $f$ denotes the energy spectrum of a fluid flow in $\R^d$
(see \S\ref{par:PMotivation} and equation~\eqref{eq:WK} below), then $\mathbb{WK}_d[f]$ is the corresponding second-order
structure function. In reference to this background that relates closely to the Wiener-Khinchin theorem
in the probabilistic analysis of time-series \cite{W30}, \cite{Kn34}, \cite{W49},
one will call $\mathbb{WK}_d[f]$  the $d$-dimensional \textsl{Wiener-Khinchin transform} of~$f$.
The names of the variables have also been chosen accordingly: in the applications, $\lambda$~stands for a length and $k$ for a spatial frequency.

\medskip
However, a physical background is not necessary to understand our results and motivations.
For instance,
one can also encounter $\mathbb{WK}_d$ when computing  the Fourier transform of radial functions in dimension $d$ 
(see \cite{Stein} or \S\ref{par:MMotivation} below) because the Hankel transform
of $k^{-d/2} f(k)$ is closely related to the Wiener-Khinchin transform once the formula is written in the following way:
\begin{equation}\label{HankelConnexion}
\Gamma\left(\tfrac{d}{2}\right) 
\int_0^\infty J_{\frac{d}{2}-1}(2\pi\lambda k) \cdot (\pi\lambda k)^{1-\frac{d}{2}} f(k) dk =
\int_0^\infty f(k) dk -\mathbb{WK}_d[f].
\end{equation}

\medskip
In this paper one will focus on the Wiener-Khinchin transform of positive functions that, roughly speaking, behave like $f(k) \simeq k^{-\alpha}$ on some interval $I=[k_1,k_2]$.
A precise definition of this similarity in terms of the slope computed in a log-log diagram is given in section~\ref{par:def}.

\medskip
Our practical goal is to find proper assumptions on the tails of $f$ outside $I$ that will ensure the existence of
an interval $J=[\lambda_1,\lambda_2]$ on which $\mathbb{WK}_d[f]$
will also be comparable to a power function with the conjugate exponent, \ie to $\lambda^{\alpha-1}$.
While this conclusion is a standard claim in physics~\cite{Frisch95}, \cite{Davidson04}, there is more to it than one
could naively expect. We are interested in \textbf{quantifying}, in the proper ranges, the similarity between the pair $(f,\mathbb{WK}_d[f])$ and the
corresponding pair of power-laws $(k^{-\alpha},\lambda^{\alpha-1})$.
For example, one would like to quantify the growth of $\lambda_2/\lambda_1$ as $k_2/k_1$ goes to $+\infty$.
Overall, our results are written in the same spirit as the error estimates in numerical analysis and rely on fine estimates of Bessel functions
and multiplicative convolutions.

\medskip
Dimensions $d=2$ and $d=3$ are obviously the most physically relevant.
But computations for a general $d\geq 2$ are overall similar, so there is little point in rejecting the highest dimensions.
Some extra work is actually required for $d\in\{2,3\}$ due to the smaller, non-integrable, decay rate of the Bessel kernel.
Dimension $d=1$ is more of an anecdote; one has $H_1(\sigma)=1-\cos(2\pi \sigma)$ thus $\mathbb{WK}_1[f]$ computes the Fourier coefficients
of the even extension of $f$ on the real line.
As the Fourier kernel does not decay but only goes to zero in a weak sense, the case $d=1$ is an exception and will be excluded when necessary.

\medskip
Our long-term goal is to consolidate the mathematical foundations of the theory of hydrodynamical turbulence.
The theory is usually based on the premise that everyone \textit{obviously} knows what it means for a function to behave
approximately like a power-law on a finite or an asymptotic range. In this article we pursue a program initiated in~\cite{FV:K41}
and show that it is perfectly possible to quantify things rigorously.
As explained in \S\ref{par:PMotivation}, there is more than one possible practical definition of homogeneous turbulence.
This article adresses the question of the equivalence between the two most prominent points of view: the spectral one (see definition~\ref{def:turbSpectral})
and the one based on $L^2$-increments (see definition~\ref{def:turbSpatial}).

\medskip
The article is structured as follows. Section \ref{par:def} contains a precise definition of the notions of local and asymptotic quasi-power-law behaviors
and notations that will be used throughout. The main results are stated and explained in \S\ref{par:results}.
This relatively long section is subdivided into six subsections that correspond to the different types of results that we have obtained:
\begin{itemize}
\item the global duality of quasi-power-laws that span the whole interval $(0,\infty)$,
\item the universal quadratic regime at the origin of Wiener-Khinchin transforms,
\item the universal constant regime at infinity of Wiener-Khinchin transforms,
\item the duality of quasi-power-laws on finite ranges,
\item some general comparison principles,
\item and a presentation of some typical numerical examples.
\end{itemize}
Results labeled ``Proposition'' are elementary and, for ease of reading, will be proved on the spot.
In \S\ref{par:MMotivation} and \S\ref{par:PMotivation}, we will then briefly expose  some of the physical background
and explain how our mathematical results translate into the language of physics.
Sections~\ref{proof:thm1}, \ref{proof:thm23} and \ref{proof:thm4} are dedicated to the proofs of the main statements.
Section~\ref{par:numerics} collects the raw data associated with the numerical study presented in~\S\ref{par:numericsexplained}.
An appendix, \S\ref{par:appendix}, recalls the various asymptotic  bounds of the Bessel functions and of the kernel $H_d$ that are useful in our proofs.

\subsection{Definitions regarding local and asymptotic power-law behavior}\label{par:def}

The graph of a (smooth) function $f:(0,+\infty)\to (0,+\infty)$ in log-log coordinates is the representation of $\log_{10} f(x)$ as a function of $\log_{10} x$.
The slope at $x=10^\xi$ is given by
\begin{equation}
\frac{d}{d\xi} \left[ \log_{10}f(10^\xi) \right] = \left.\frac{x f'(x)}{f(x)}\right|_{x=10^\xi}.
\end{equation}
This representation is central in engineering and in most lab work regarding fluid turbulence because of the following property:
\textit{a function $f$ is a power-law \ie $f(x)=c x^\alpha$ if and only if its log-log representation is a straight line of slope $\alpha$}.
The goal of this section is to quantify the similarity between a given function and a power-law, on a specified range.

\medskip
Let us introduce the following three quantities called the \textbf{power-law gauges} of $f$.
\begin{dfn}
For $\alpha\in\R$ and $a<b$ in $[0,+\infty]$, one defines an integral power-law gauge:
\begin{equation}
\|f\|_{\power_1^\alpha(a,b)}=\int_a^b \left| \frac{x f'(x)}{f(x)} - \alpha \right| \frac{dx}{x}
\end{equation}
and a pointwise one  (see \cite{FV:K41}) by:
\begin{equation}
\|f\|_{\power^\alpha_\infty (a,b)} =  \sup_{x\in[a,b]} \left| \frac{x f'(x)}{f(x)} - \alpha \right|.
\end{equation}
Instead of considering $\|f\|_{\power_1^\alpha(a,b)}$,
one sometimes has  to ``relax'' the absolute values outside the integral, the tradeoff being that one then has to check all possible sub-intervals.
One defines:
\begin{equation}\label{defG0}
\|f\|_{\power_0^\alpha(a,b)}= \sup_{a',b'\in(a,b)}\left|\int_{a'}^{b'} \left( \frac{x f'(x)}{f(x)} - \alpha \right) \frac{dx}{x}\right|
= \sup_{a',b'\in(a,b)}\left| \left[ \log \left(\frac{f(x)}{x^\alpha}\right)\right]_{a'}^{b'} \right|
\end{equation}
with the notation $[g(x)]_{a'}^{b'} = g(b')-g(a')$.
\end{dfn}
Let us point out that the three power-law gauges are homogeneous of degree 0 (see \eqref{homogeneousGauge} below) and that:
\begin{equation}\label{gauges_inequality}
\|f\|_{\power_0^\alpha(a,b)} \leq \|f\|_{\power_1^\alpha(a,b)}\leq \|f\|_{\power^\alpha_\infty (a,b)} \cdot \log\left(\frac{b}{a}\right).
\end{equation}
If one defines $\varepsilon : (0,+\infty)\to\R$ by the ODE $x f'(x) = (\alpha+\varepsilon(x)) f(x)$, then
\begin{equation}\label{ODEdefGauges}
\|\varepsilon\|_{L^1(a,b;\frac{dx}{x})}
= \|f\|_{ \power^\alpha_1(a,b) } \quad\text{and}\quad
\|\varepsilon\|_{L^\infty(a,b)} = \|f\|_{ \power^\alpha_\infty(a,b) }.
\end{equation}

\medskip
In this article, we will need all three gauges; it is not always possible to capture $\power_\infty^\alpha$ in an optimal way while the integral gauges
$\power_1^\alpha$ and $\power_0^\alpha$ offer more flexibility. Note that defining the gauge $\power_0^\alpha$
does not require as much smoothness on $f$; one could for example define it provided only that $\log f\in L^\infty_\text{loc}$
instead of asking that \eg $f$ is a non-vanishing function in $W^{1,\infty}_{\text{loc}}$.
While defining intermediary $L^p$-based gauges $\power_p^\alpha$ is perfectly possible for $1<p<\infty$, it is not necessary for what follows.

\paragraph{Geometrical intrepretation of $\power_\infty^\alpha$.} 
The gauge $\|f\|_{\power^\alpha_\infty (a,b)}$ is the maximal deviation from $\alpha$ of the slope of the graph of $f$ in log-log coordinates.
However, if $b/a\gg1$, its smallness is not enough to ensure that the graph remains close to a straight line of slope $\alpha$ (see the first two examples below).

\paragraph{Geometrical intrepretation of $\power_0^\alpha$ and  $\power_1^\alpha$.} 
On the contrary, the geometrical meaning of $\|f\|_{\power_0^\alpha(a,b)}$ and $\|f\|_{\power_1^\alpha(a,b)}$ is a
measure of the similarity between the graph of $f$ in log-log coordinates and a straight line of slope $\alpha$.
In the case of a finite interval, the smallness of $\|f\|_{\power_0^\alpha(a,b)}$ or $\|f\|_{\power_1^\alpha(a,b)}$ ensures
that the log-log graph remains close to a straight line on that interval. If $a=0$ or $b=\infty$, it ensures
the existence of an asymptote in log-log coordinates.

\medskip\noindent
In~\cite{FV:K41}, the quantity $ \|f\|_{\power^\alpha_\infty (a,b)} \log\left(\frac{b}{a}\right)$ was used instead,
and according to~\eqref{gauges_inequality}, it serves the same geometrical purpose as $\|f\|_{\power_0^\alpha(a,b)}$ or $\|f\|_{\power_1^\alpha(a,b)}$.

\paragraph{Examples.}
The following examples illustrate how the different gauges catch the various behaviors that one can expect in log-log coordinates.
\begin{enumerate}
\item
The function $f(x)=c_0 x^{\alpha'}$ satisfies
\[
\|f\|_{\power_\infty^{\alpha}(a,b)} = |\alpha-\alpha'| 
\quad\text{and}\quad
\|f\|_{\power_1^{\alpha}(a,b)} = |\alpha-\alpha'|  \cdot \log\left(\frac{b}{a}\right).
\]
In log-log coordinates, the best uniform approximation of the graph of $f$ by a line of slope $\alpha$ is 
given by the function $\tilde{f}(x)=c_0 (ab)^{\frac{\alpha'-\alpha}{2}} x^\alpha$.
The geometric deviation between the two graphs is:
\[
\|\log \tilde{f} - \log f\|_{L^\infty(a,b)}= \frac{1}{2}\|f\|_{\power_1^{\alpha}(a,b)}.
\]
The smallness of any of the three quantities that appear in~\eqref{gauges_inequality} thus has a global geometrical meaning, while the smallness of the 
gauge $\power_\infty^{\alpha}(a,b)$ by itself (without a $\log(b/a)$ factor) does not.
\item 
In log-log coordinates, the function$f(x)=c_0x^\alpha \log x$ has an asymptotic direction $\xi\mapsto \alpha\xi$, but no asymptote.
The gauge $\|f\|_{\power^\alpha_\infty(a,\infty)}=\frac{1}{\log a}\underset{a\to+\infty}{\longrightarrow}0$ captures the asymptotic direction,
while the gauge $\|f\|_{\power_1^\alpha(a,\infty)}=\int_a^\infty \frac{dx}{x\log x}=+\infty$ quantifies the lack of an asymptote.
\item 
On the contrary, for $\beta>1$, the function $f(x)=c_0 x^{\alpha} \exp(-\frac{1}{(\beta-1)\log^{\beta-1}(x)})$ satisfies
\[
\|f\|_{\power_1^\alpha(a,\infty)}
= \int_a^\infty \frac{dx}{x \log^\beta x}=\frac{1}{(\beta-1)\log^{\beta-1} a}\underset{a\to+\infty}{\longrightarrow}0
\]
and its graph is indeed asymptotic to the line $\xi\mapsto \alpha\xi+\log c_0$ in log-log coordinates. This example illustrates again that
the gauge $\power_1^{\alpha}(a,b)$ correctly captures the geometry of the log-log graph.
\end{enumerate}

\pagebreak[2]\bigskip
The two following definitions are central for this article and match up (as closely as possible)
with the common language and notations used in physics and engineering.
\begin{dfn}\label{HFR}
Given $a<b$ in $[0,+\infty]$, $\alpha\in\R$ and $\varepsilon_0>0$,
a smooth positive function $f$ on $\R_+^\ast$
is said to be a \textbf{quasi-power-law of exponent $\alpha$ on~$[a,b]$, up to the tolerance $\varepsilon_0$} if
\begin{equation}\label{power_law}
\|f\|_{\power_0^\alpha(a,b)} \leq \varepsilon_0.
\end{equation}
One will denote this property by $f(x) \overset{\varepsilon_0}{\underset{[a,b]}{\propto}} x^\alpha$
or $f(x)\propto x^\alpha$ if $a,b$ and $\varepsilon_0$ are clear from the context.
\end{dfn}
\begin{dfn}
When not all parameters are specified, one says the following.
\begin{itemize}
\item $f$ is a \textbf{quasi-power-law on a finite range $[a,b]$} if there exists $\alpha\in\R$ and a tolerance~$\varepsilon_0\ll1$ 
such that 
\begin{equation}
f(x) \overset{\varepsilon_0}{\underset{a,b}{\propto}} x^\alpha.
\end{equation}
\item $f$ is \textbf{asymptotically homogeneous near 0} (resp. \textbf{at infinity}) if there exists $\alpha\in\R$ such that
\begin{equation}
\lim_{\substack{x\to 0\\\text{resp. }x\to +\infty}}\log \left(\frac{f(x)}{x^\alpha}\right) \in \R.
\end{equation}
In this case, for any $\varepsilon>0$, one can find $b$ (resp. $a$) such that 
$f(x) \overset{\varepsilon}{\underset{0,b}{\propto}} x^\alpha$ (resp. $f(x) \overset{\varepsilon}{\underset{a,\infty}{\propto}} x^\alpha$).
\end{itemize}
\end{dfn}
\paragraph{Remarks.}
\begin{itemize}
\item
It is tempting to describe quasi-power-laws as \textit{quasi-homogeneous functions};  however, this expression is already in use in some
branches of mathematics and mathematical physics\footnote{see \eg\url{https://en.wikipedia.org/wiki/Quasi-homogeneous_polynomial}.}
and is associated with anisotropic scalings.
In order to avoid an unnecessary confusion, we will \textbf{not} use it.
\item
In  common language and in physics textbooks, the threshold of smallness for $\varepsilon_0$ is left
to the common sense of the reader and can, for example, be expected to be of the same order of magnitude
as the error estimates on the experimental or numerical data.
\end{itemize}

\bigskip
By compactness, any smooth positive function satisfies \eqref{power_law} on any finite interval $[a,b]$ but for some ridiculously large $\varepsilon_0$.
The significant question is thus to decide whether it is possible to have $\varepsilon_0\ll1$, which might, in turn, require some compromises in the choice of
the values of $a$ and $b$.
On a finite interval, one cannot expect in general to shrink $\varepsilon_0$ to an arbitrary small value because 
any function $f$ that satisfies~\eqref{power_law} with $\varepsilon_0=0$ is exactly of the form $f(x)=cx^\alpha$ on $[a,b]$ for some constant $c\in\R$.

\begin{prop}\label{prop:powerlawconsequence}
If a smooth positive function $f$ is a quasi-power-law of exponent $\alpha$ on $[a,b]$, up to a given tolerance~$\varepsilon_0$,
 one can find $c_0>0$ such that  in log-log coordinates,
 the graph of $f$ does not deviate from the line $\xi \mapsto \alpha \xi +\log c_0$ by more than $\pm\varepsilon_0$, which is to say, one has:
 \begin{equation}
c_0 x^\alpha e^{-\varepsilon_0} \leq  f(x)\leq c_0 x^\alpha e^{\varepsilon_0}.
 \end{equation}
Moreover, if $\varepsilon_0<1$, the relative error committed in replacing $f(x)$ by the power-law $c_0x^\alpha$ is of order~$\varepsilon_0$:
\begin{equation}\label{powerlawconsequence}
\sup_{x\in[a,b]} \left|\frac{f(x)-c_0 x^\alpha}{f(x)}\right| < 2\varepsilon_0.
\end{equation}
\end{prop}
\Proof As the proof is short and enlightening, it is worth not postponing it.
Let us choose $b_0\in[a,b]$ and define $c_0=f(b_0)/b_0^\alpha$. One has, for any $x\in[a,b]$:
\[
\log c_0 = \log f(x) - \alpha \log x + \int_{x}^{b_0} \left( \frac{t f'(t)}{f(t)} - \alpha \right) \frac{dt}{t}\cdotp
\]
For each choice of $b_0$, the function
\[
\theta(x)=-\int_{x}^{b_0} \left( \frac{t f'(t)}{f(t)} - \alpha \right) \frac{dt}{t}
\]
satisfies $f(x)=c_0x^\alpha e^{\theta(x)}$ and $\|\theta\|_{L^\infty(a,b)}\leq \|f\|_{\power_0^\alpha(a,b)} \leq \varepsilon_0$. 
In particular, the vertical separation between the line $\xi \mapsto \alpha \xi +\log c_0$ and the graph of $\xi\mapsto \log f(e^\xi)$, which is $|\theta(e^\xi)|$,
does not exceed $\varepsilon_0$.
Moreover, for $\varepsilon_0<1$, one has:
\[
\left|\frac{f(x)-c_0 x^\alpha}{f(x)}\right| = |1-e^{-\theta(x)}|  \leq e^{|\theta(x)|}-1 \leq e^{\varepsilon_0}-1 < 2\varepsilon_0
\]
for any $x\in[a,b]$.\cqfd

\paragraph{Remark.} If one assumes instead that $\|f\|_{\power_1^\alpha(a,b)}\leq \varepsilon_0$
then $f$ is not only a quasi-power-law thanks to~\eqref{gauges_inequality},
but $x f'(x)/f(x) - \alpha$ is also integrable on $[a,b]$ so $\log(f(x)/x^\alpha)$ has a finite limit at the boundaries.
This observation is especially useful when $a=0$ or $b=\infty$.
\par

\bigskip\noindent
The power-law gauges are not norms or semi-norms, but they satisfy the following properties.
\begin{prop}
\begin{subequations}
For $a<b$,  $\alpha,\beta\in\R$ and any $f,g:(0,+\infty)\to(0,+\infty)$, one has
the following properties, for both the gauges $\power_1^\alpha(a,b)$ and $\power^\alpha_\infty(a,b)$:
\begin{gather}
\|f\|_{\power_i^\alpha(a,b)} = \|x^{-\alpha} f(x)\|_{\power_i^0(a,b)},\\
\label{homogeneousGauge}
\forall r>0,\qquad \|f\|_{\power_i^\alpha(a,b)} = \|r f\|_{\power_i^\alpha(a,b)},\\
\|f+g\|_{\power_i^\alpha(a,b)} \leq \|f\|_{\power_i^\alpha(a,b)}+\|g\|_{\power_i^\alpha(a,b)},\\
\|fg\|_{\power_i^{\alpha+\beta}(a,b)} \leq \|f\|_{\power_i^{\alpha}(a,b)}+\|g\|_{\power_i^\beta(a,b)}.
\end{gather}
Finally, if $f>g$ on $[a,b]$, one also has:
\begin{equation}
\|f-g\|_{\power_i^\alpha(a,b)} \leq \|f\|_{\power_i^\alpha(a,b)} \left\|\frac{f}{f-g}\right\|_{L^\infty(a,b)}
+\|g\|_{\power_i^\alpha(a,b)}\left\|\frac{g}{f-g}\right\|_{L^\infty(a,b)}.
\end{equation}
\end{subequations}
\end{prop}

\subsection{Statement of the main results}\label{par:results}

For $1<\alpha<3$, the Wiener-Khinchin transform of $f(k)=k^{-\alpha}$ is also a homogeneous function:
\begin{equation}
\mathbb{WK}_d[k^{-\alpha}](\lambda) = c_d(\alpha) \lambda^{\alpha-1}
\end{equation}
with, for example, when $d=2$ or $d=3$:
\[
c_2(\alpha)=\frac{\pi^{\alpha-1}}{2}\frac{|\Gamma\left(-\frac{\alpha-1}{2}\right)|}{\Gamma\left(\frac{\alpha+1}{2}\right)}
\qquad\text{and}\qquad
c_3(\alpha)=\begin{cases}
(2\pi)^{\alpha-1}\Gamma(-\alpha)\sin\left(\frac{\alpha\pi}{2}\right) & \text{if }\alpha\neq 2,\\
\frac{\pi^2}{2}& \text{if }\alpha = 2.
\end{cases}
\]
The restriction $\alpha\in(1,3)$ is natural, regardless of the dimension $d$, because for $\alpha\leq 1$ the integral formula~\eqref{main}
diverges at infinity, while for $\alpha\geq 3$ the integral formula
diverges at the origin. Let us point out that the applications to hydrodynamics (see \S\ref{par:PMotivation}) all fit in this range.

\subsubsection{Global duality of quasi-power-laws}

Our first result provides a quantitative estimate of global stability around the pair $(k^{-\alpha} ; c_d(\alpha) \lambda^{\alpha-1})$.
\begin{thm}\label{thm1}
For $1<\alpha<3$ and $d\geq2$, the following pointwise inequality
\begin{equation}\label{global1}
\|\mathbb{WK}_d[f]\|_{\power^{\alpha-1}_\infty(0,\infty)} \leq \|f\|_{\power^{-\alpha}_\infty(0,\infty)}
\end{equation}
holds for any smooth function $f:\R_+^\ast\to\R_+^\ast$.
For the integral power-law gauges, one has:
\begin{equation}\label{global2}
\|\mathbb{WK}_d[f]\|_{\power_1^{\alpha-1}(0,\infty)} \leq \frac{c_d^+}{c_d^-} 
\|f\|_{\power_1^{-\alpha}(0,\infty)} \exp\left(2\|f\|_{\power_0^{-\alpha}(0,\infty)}\right).
\end{equation}
The constants $c_d^\pm$ are defined by~\eqref{kernel:intermidiary} in the appendix and computed in Table~\ref{Table:cpm}.
\end{thm}

The pointwise statement indicates that if the fluctuations of the log-log slope of~$f$ around $-\alpha$ are bounded, then
the fluctuations of the log-log slope of $\mathbb{WK}_d[f]$ around $\alpha-1$ will obey the exact same bounds.
The second statement quantifies instead the fact that the global similarity with a straight line is inherited through the Wiener-Khinchin transform.
Note that thanks to~\eqref{gauges_inequality}, the right-hand side of~\eqref{global2} becomes linear in $\|f\|_{\power_1^{-\alpha}(0,\infty)}$ when this gauge is small enough.
Note also that both estimates are uniform with respect to $\alpha\in(1,3)$.
This result is illustrated on figure~\ref{Fig:global} below.

\bigskip
For the rest of this article, we will focus on functions $f$ whose tails near 0 and at infinity have no reason to match $c_0k^{-\alpha}$.
In this case, theorem~\ref{thm1} still holds but does not provide any useful insight.
A typical example is functions that are quasi-power-laws on only a finite range.

\paragraph{Example.}
Theorem~\ref{thm1} is illustrated for $d=2$ and $d=3$ on figure~\ref{Fig:global}.
The function $f$ is obtained by solving $\frac{k f'(k)}{f(k)}+\frac{3}{2}=50\sin(5t)\exp\left(-t-\frac{1}{t}\right)$.
One has $\|f\|_{\power_0^{-3/2}(0,\infty)}\simeq 3.16$
and $\|f\|_{\power_1^{-3/2}(0,\infty)}\simeq 7.27$ and $\|f\|_{\power_\infty^{-3/2}(0,\infty)}\simeq 6.75$.
The maximum slopes can be read in the insets drawings; one can check that the inequality \eqref{global1} is satisfied. 
The fact that $\log\lambda \mapsto \log \mathbb{WK}_d[f](\lambda)$ is overall ``close'' to a straight line of
slope~$\alpha-1$ illustrates \eqref{global2}.

\begin{figure}[H]
\captionsetup{width=.9\linewidth}
\begin{center}
\includegraphics[width=.9\textwidth]{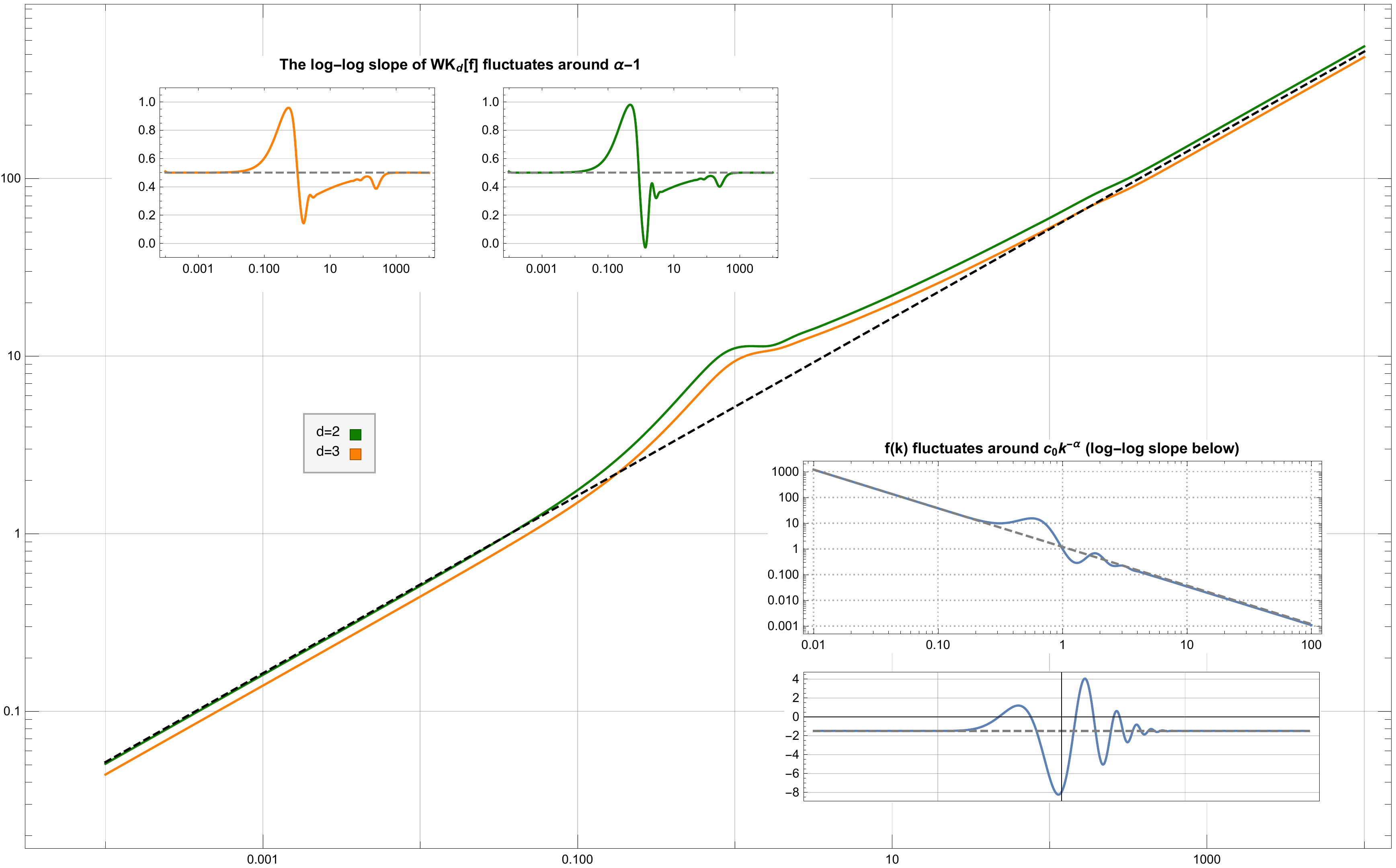}\\
\caption{\label{Fig:global}\small
Plot of $\mathbb{WK}_d(f)$ with $d=2$ (green), $d=3$ (orange) and a reference $c_1\lambda^{\alpha-1}$
for a function $f$ (bottom right, top inset) that fluctuates around $c_0 k^{-\alpha}$.
The pointwise log-log slopes of $f$ (bottom right, lower inset) and of $\mathbb{WK}_d(f)$ (top left insets) are also given.
}
\end{center}
\end{figure}

\subsubsection{Universal quadratic regime at the origin}

Our second result states the existence of a universal regime $\mathbb{WK}_d[f](\lambda)\propto\lambda^2$ as $\lambda\to0^+$.
More precisely, one can show that the $\power^2_\infty(0,b)$ and $\power^2_1(0,b)$ gauges of $\mathbb{WK}_d[f]$ go to zero when $b\to0$ and compute an exact decay rate.
\begin{thm}\label{thm2}
Let us consider $d\geq1$ and a smooth positive function $f$ such that
\begin{equation}
\label{smallscale:assumption}
\int_0^\infty (1+k^2) f(k) dk <\infty.
\end{equation}
The function $\mathbb{WK}_d[f](\lambda)$ is asymptotically homogeneous to $\lambda^2$ at the origin.
Precisely, let us define
\begin{equation}
K_\delta^\sharp = \inf \left\{ K>0 \,\Big\vert\,
K^2 \int_K^\infty f(k) dk \leq \frac{\delta^4 }{2} \int_0^K k^2 f(k) dk
\right\}.
\end{equation}
In dimension $d\in\{1,2\}$, the definition of $K_\delta^\sharp$ should be amended to ensure also that: 
\begin{equation}
\forall K\geq K_\delta^\sharp, \qquad 
\int_K^\infty k^2 f(k) dk \leq 
\frac{\delta^2}{2} \int_0^K k^2 f(k) dk.
\end{equation}
Then for any $\delta\leq \delta_0=\sqrt{d/(2\pi^2)}$, one has:
\begin{equation}\label{smallscale:conclusion1}
\|\mathbb{WK}_d[f]\|_{\power^2_\infty\big(0,\frac{\delta}{K_\delta^\sharp}\big)} \leq 
\frac{3 d C_L}{2\pi^2} \, \delta^2
\end{equation}
where $C_L$ is a numerical constant given in the appendix (Lemma~\ref{lemma:L} and Table~\ref{T:CL}).
Moreover, if $f$ has a finite fourth-order momentum or if $K_\delta^\sharp$ has a bounded log-log slope as $\delta\to0$,
then:
\begin{equation}\label{smallscale:conclusion2}
\| \mathbb{WK}_d[f]\|_{\power_1^2\big(0,\frac{\delta}{K_{\delta}^\sharp}\big)}
\leq C_f \, \delta^2
\end{equation}
with a constant $C_f$ that depends on $f$ through \eqref{UR0C1}-\eqref{UR0C2}.
When $d\geq3$, the log-log slope of $\mathbb{WK}_d[f](\lambda)$ never exceeds $2$, \ie:
\begin{equation}\label{upperslope}
\forall \lambda>0,\qquad \frac{\lambda \mathbb{WK}_d[f]'(\lambda)}{\mathbb{WK}_d[f](\lambda)} < 2.
\end{equation}
In particular, the graph of $\mathbb{WK}_d[f](\lambda)$ stays below $c_d(\int_0^\infty k^2 f(k)dk) \lambda^2$ 
with $c_d=\frac{d}{2\pi^2}+\frac{2\pi^2}{d}$ and the vertical distance between the two log-log graphs increases with $\lambda$ on the whole interval $(0,\infty)$.
\end{thm}

\medskip
The existence of $K_\delta^\sharp$ is ensured by~\eqref{smallscale:assumption} and one usually expects $K_\delta^\sharp$  to be ``large''.
The function~$\delta/K_\delta^\sharp$ is an increasing  sub-linear function of~$\delta$.
In particular, $\delta_0/K_{\delta_0}^\sharp$ is the largest scale at which one can expect to
see $\mathbb{WK}_d[f](\lambda)\propto\lambda^2$.
Estimates~\eqref{smallscale:conclusion1} and~\eqref{smallscale:conclusion2} state that
to improve the tolerance by  a numerical factor~$10^{-p}$,
one will need to restrict this maximal scale by a factor~$10^{-p/2}$ and
possibly much more (as described by $K_\delta^\sharp$) if $f$ has a heavy tail.

\paragraph{Remark.}
For $d\geq3$,
the increasing distance between $\mathbb{WK}_d[f](\lambda)$ and its asymptote does not quite imply the concavity
of the graph of $\mathbb{WK}_d[f](\lambda)$ in log-log scale (see the last example of \S\ref{par:numericsexplained}), but
it is a strong geometrical constraint, especially as $\lambda\to0^+$.
\par

\paragraph{Examples.}
In the physical applications, $\delta$ is dimensionless, $K_\delta^\sharp$ has the dimension
of the variable $k$ in $f(k)$ and~$\delta/K_\delta^\sharp$ has the dimension of the variable $\lambda\sim k^{-1}$ in $\mathbb{WK}_d[f](\lambda)$.
With the applications in mind, let us compute the asymptotic behavior of $K_\delta^\sharp$ when $\delta\to0$. 
For simplicity,  we will assume that the dimension $d\geq3$ and that $K_\delta^\sharp$ is large enough so that, at the leading order:
\[
\int_0^{K_\delta^\sharp} k^2 f(k)dk \simeq \int_0^\infty k^2 f(k)dk.
\]
\begin{enumerate}
\item
If $f$ is compactly supported on $[0,K_0]$, then $K_\delta^\sharp\simeq K_0$ and $\delta/K_\delta^\sharp\propto \delta$; therefore, one has:
\[
\|\mathbb{WK}_d[f]\|_{\power^2_1(0,\delta)} = \mathcal{O}(\delta^2).
\]
\item
If the tail of $f$ obeys an exponential law $f(k)\sim c_0 e^{-\lambda_0 k}$, then the equation that defines $K_\delta^\sharp$ reads
\[
\frac{c_0}{\lambda_0}  (K_\delta^\sharp)^2 e^{-\lambda_0 K_\delta^\sharp} \simeq \frac{\delta^4}{2} \int_0^\infty k^2 f(k) dk
\]
which roughly means that $K_\delta^\sharp \propto |\log\delta|$, \ie
\[
\|\mathbb{WK}_d[f]\|_{\power^2_1\big(0,\frac{\delta}{|\log \delta|}\big)} =\mathcal{O}(\delta^2).
\]
\item
If the tail of $f$ obeys a power-law $f(k)\sim c_0 k^{-\beta}$ at infinity with $\beta>3$, one gets
 $K_\delta^\sharp \propto \delta^{-4/(\beta-3)}$ and thus:
\[
\|\mathbb{WK}_d[f]\|_{\power^2_1\big(0,\delta^\frac{\beta+1}{\beta-3}\big)}=\mathcal{O}(\delta^2)
\qquad\ie\qquad
\|\mathbb{WK}_d[f]\|_{\power^2_1(0,\delta)}=\mathcal{O}(\delta^\frac{2(\beta-3)}{\beta+1}).
\]
If $\beta=3$, one applies \eqref{UR0AltAss} to get  $K_\delta^\sharp \propto \exp(\delta^{-4})$ and
 $\|\mathbb{WK}_d[f]\|_{\power^2_1\big(0,\delta \exp(-\delta^{-4})\big)}=\mathcal{O}(\delta^2)$.
\end{enumerate}
The extent of the range of the universal quadratic regime of $\mathbb{WK}_d[f]$ thus provides an indirect but computable insight into the tail of $f$ at infinity.

\subsubsection{Universal constant regime at infinity}

Our third result states the existence of an asymptotically constant
regime $\mathbb{WK}_d[f](\lambda)\propto1$ as $\lambda\to+\infty$.

\begin{thm}\label{thm3}
For any $d\geq2$ and any smooth positive function $f$ such that
\begin{equation}
\int_0^\infty f(k) dk < \infty,
\end{equation}
the function $\mathbb{WK}_d[f]$ satisfies
\begin{equation}
\lim_{\lambda\to+\infty}\mathbb{WK}_d[f](\lambda) = \int_0^\infty f(k) dk
\end{equation}
and is thus asymptotically homogeneous to a constant at infinity.
More precisely, for $\eta\geq \eta_0$ defined by~\eqref{defmu0}, one has:
\begin{equation}\label{thm3C1}
\|\mathbb{WK}_d[f]\|_{\power^0_\infty(\frac{\eta}{K_\eta^\flat},\infty)} \leq
\begin{cases}
C_d\, \eta^{-\frac{d-3}{2}} & \text{if } d\geq4,\\
C_d\, \eta^{-\frac{d-1}{2}} & \text{if }d\in \{ 2,3 \}
\end{cases}
\end{equation}
with
\begin{equation}
K_\eta^\flat = \sup \left\{ K>0 \,\Big\vert\,
\int_0^K f(k) dk \leq \frac{1}{2} \: \eta^{-\frac{d-1}{2}} \int_K^\infty f(k) dk.
\right\}
\end{equation}
and a constant $C_d$ that only depends on $d$.
In dimension $d\in\{2,3\}$, the definition of $K_\eta^\flat$ should be amended to ensure that: 
\begin{align}\label{thm3:Ad3}
\sup_{\lambda\geq \mu / K_\eta^\flat} \left|\int_0
^\infty e^{2 i \pi \lambda k} f(k) dk \right|
&\leq \eta^{-1} \int_0^\infty f(k) dk
\qquad (d=3),\\ \label{thm3:Ad2}
\sup_{\lambda\geq \mu / K_\eta^\flat} \left|\int_0
^\infty \lambda k J_1(2 \pi \lambda k)  f(k) dk \right|
& \leq \eta^{-1/2} \int_0^\infty f(k) dk
\qquad (d=2).
\end{align}
Moreover, for $d\geq 3$, the log-log slope of $\mathbb{WK}_d[f](\lambda)$ is globally bounded between $-2$ and $+2$:
\begin{equation}
\|\mathbb{WK}_d[f]\|_{\power^0_\infty(0,\infty)} \leq2.
\end{equation}
\end{thm}

\paragraph{Remarks.}
\begin{itemize}
\item
For the applications, note that $\eta$ is dimensionless, while $K_\eta^\flat$ has (like $K_\delta^\sharp$ before) the dimension of the variable $k$
in the expression of $f(k)$ or that of $\lambda^{-1}$  in $\mathbb{WK}_d[f](\lambda)$.
\item
If one assumes some additional regularity and decay on $f$, one can better~\eqref{thm3C1} by one order of magnitude when $d\geq 2$
and without even assuming~\eqref{thm3:Ad3}-\eqref{thm3:Ad2}:
\begin{equation}\label{thm3C1bis}
\|\mathbb{WK}_d[f]\|_{\power^0_\infty(\frac{\eta}{K_\eta^\flat},\infty)} 
\leq C  \left(
\frac{\int_0^\infty |f'(k)| k^{-\frac{d-3}{2}}dk}{\int_0^\infty f(k) dk}
\right) \eta^{-\frac{d-1}{2}}  \qquad (d\geq 2).
\end{equation}
This alternate technique also provides an estimate in dimension $d=1$:
\begin{equation}
\|\mathbb{WK}_1[f]\|_{\power^0_\infty(\frac{\eta}{K_\eta^\flat},\infty)} 
\leq C  \left(
\frac{\int_0^\infty |f''(k)| k dk}{\int_0^\infty f(k) dk}
\right) \eta^{-1}
\qquad (d=1).
\end{equation}
In general, one does not expect a better decay rate of $\power^0_\infty$ than this improved one.

\item For $d\geq2$, one can also control the $\power^0_1$ gauge by:
\begin{equation}
\| \mathbb{WK}_d[f]\|_{\power_1^0(\frac{\eta}{K_\eta^\flat},\infty)} \leq C \left(
\frac{\int_0^\infty |f'(k)| k^{-\frac{d-3}{2}}dk}{\int_0^\infty f(k) dk}
\right)
\int_{\eta}^\infty \left(1-\frac{s (K_s^\flat)'}{K_s^\flat}\right) \frac{ds}{s^{\frac{d+1}{2}}}
\end{equation}
which is $\mathcal{O}(\eta^{-\frac{d-1}{2}})$ if \eg $K_\eta^\flat$ has a bounded log-log slope as $\eta\to\infty$
(see the final remark of \S\ref{par:URinf}).
\item
One usually expects $K_\eta^\flat$ to be ``small'';
precisely, one has $K_\eta^\flat < K_\delta^\sharp$ for any $\delta\leq\delta_0$ and $\eta\geq \eta_0$
where $\delta_0$ and $\mu_0$ are the respective thresholds in theorems~\ref{thm2} and~\ref{thm3}.
Inequality~\eqref{KflatKSharp} legitimates our choice of musical notations~$\flat$ and~$\sharp$.
It also ensures that there is ``room'' for an intermediary range in-between the two universal asymptotic regimes.
This tempered range is our next focus.
\end{itemize}
\par

\subsubsection{Duality of quasi-power-laws on finite ranges}

The fourth result is of particular relevance for the applications.
It states that $\mathbb{WK}_d[f]$ is a quasi-power-law on a finite range as soon as $f$ is too,
provided that the tails at zero and infinity are not too heavy. Moreover, the two ranges
are, as dimensional analysis suggests, roughly the inverse of each other.
\begin{thm}\label{thm4}
Let $d\geq3$ and $C_1, C_2>0$. For any smooth positive function $f$ and real numbers $k_1<k_2$ such that
\begin{equation}\label{assumption0:tail}
\int_0^{k_1} k^2f(k) dk \leq C_1 k_1^3 f(k_1)
\qquad\text{and}\qquad
\int_{k_2}^\infty f(k)dk  \leq C_2 k_2 f(k_2),
\end{equation}
one has, for $1<\alpha<3$:
\begin{equation}\label{midrange:conclusion}
\|\mathbb{WK}_d[f]\|_{\power^{\alpha-1}_\infty(\frac{\mu}{k_2},\frac{\varepsilon}{k_1})} 
\leq \|f\|_{\power^{-\alpha}_\infty(k_1,k_2)}
+ C \exp\left(\|f\|_{\power^{-\alpha}_{0}(k_1,k_2)}\right) \sigma_\alpha(\varepsilon,\mu)
\end{equation}
with a constant $C$ whose dependence on $C_1$, $C_2$, $d$ and $\alpha$ is given by~\eqref{HRconstDimd}.
The arbitrary parameters~$\varepsilon$~(small) and~$\mu$~(large) are such that:
\begin{equation}\label{epssmallmularge0}
\sigma_\alpha(\varepsilon,\mu) := (\alpha-1) (\pi\varepsilon)^{3-\alpha} + (3-\alpha) (\pi \mu)^{-(\alpha-1)} \leq 1
\end{equation}
and $\frac{k_2}{k_1} > \frac{\mu}{\varepsilon}$.
Thanks to~\eqref{gauges_inequality}, one also has:
\begin{equation}
\|\mathbb{WK}_d[f]\|_{\power^{\alpha-1}_1(\frac{\mu}{k_2},\frac{\varepsilon}{k_1})} 
\leq \|\mathbb{WK}_d[f]\|_{\power^{\alpha-1}_\infty(\frac{\mu}{k_2},\frac{\varepsilon}{k_1})} 
\log\left( \frac{k_2}{k_1} \frac{\varepsilon}{\mu}\right).
\end{equation}
In dimensions $d\in\{1,2\}$ the same result holds if one requires additionally that:
\begin{equation}\label{assumption0:dim2}
\int_{k_2}^\infty |f'(k)| kdk 
\leq C_2 k_2 f(k_2)
\end{equation}
and $C$ is then given by~\eqref{HRconstDim2}-\eqref{HRconstDim1}.
\end{thm}

\begin{figure}[H]
\captionsetup{width=300pt}
\begin{center}
\includegraphics[width=250pt]{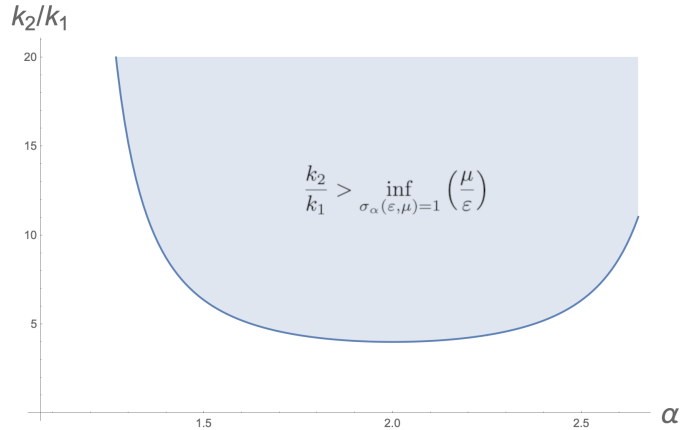}
\caption{\label{fig:reynolds}\small
Minimal size of the range on which $f$ has to be a quasi-power-law for theorem~\ref{thm4} to hold.
For the applications to turbulence (see \S\ref{par:PMotivation}), this is a minimal Reynolds number;
a practical threshold should be at least one or two orders of magnitude above.}
\end{center}
\end{figure}

In theorem~\ref{thm4}, one can always attempt to lower the value of $\varepsilon$ and increase the value of $\mu$, which then shrinks
the interval $(\frac{\mu}{k_2},\frac{\varepsilon}{k_1})$, but the restriction $\frac{k_2}{k_1} > \frac{\mu}{\varepsilon}$  prevents it from vanishing.
Contrary to what happens in theorems~\ref{thm2} and~\ref{thm3}, adjusting the values of $\varepsilon,\mu$ will not arbitrarily shrink 
the right-hand side of~\eqref{midrange:conclusion}.
Note also that the restrictions on $\varepsilon$ and $\mu$ cannot all be satisfied unless $k_2/k_1$ exceeds some minimal value
shown on figure~\ref{fig:reynolds}. In other words, one cannot guaranty that $\mathbb{WK}_d[f]$ is a quasi-power-law on some range
unless $f$ is a quasi-power-law on a large enough dual range.

Numerically (see \S\ref{par:numericsexplained} and \S\ref{par:numerics}), it seems that one looses roughly one decade
of quasi-power-law range when the $\mathbb{WK}_d$ transform is applied.

The optimal choice for $\varepsilon$ and $\mu$ is given implicitly by the equation
\begin{equation}\label{optmuepschoice}
\sigma_\alpha(\varepsilon,\mu) \simeq \min\left\{ 1;C_{d,\alpha} \, 
\frac{  \|f\|_{\power^{-\alpha}_\infty(k_1,k_2)} 
 \exp\left(- \|f\|_{\power^{-\alpha}_0(k_1,k_2)}\right)
}{ 
1+\frac{ \int_0^{k_1} k^2 f(k) dk }{k_1^3 f(k_1)} + \frac{\int_{k_2}^\infty f(k) dk}{k_2 f(k_2)}
}
\right\}
\end{equation}
and along that curve in the $(\varepsilon,\mu)$-plane, one can look for the values that minimize $\mu/\varepsilon$.
Theorem~\ref{thm4} is applicable if this infimum does not exceed $k_2/k_1$.
Using~\eqref{gauges_inequality} one could also replace $\exp(-\|f\|_{\power^{-\alpha}_0(k_1,k_2)})$ 
in the previous formula by the power $\left(k_2/k_1\right)^{-\|f\|_{\power^{-\alpha}_\infty(k_1,k_2)}}$.

\begin{cor} If $d\geq3$ and $f$ is a smooth positive function such that
\[
\int_0^\infty (1+k^2) f(k) dk <\infty,
\]
then $\|\mathbb{WK}_d[f]\|_{\power^{\alpha-1}_\infty(\frac{\mu}{k_2},\frac{\varepsilon}{k_1})} 
\leq 2 \|f\|_{\power^{-\alpha}_\infty(k_1,k_2)}$
for any $k_1$, $k_2$, $\varepsilon$, $\mu$ that satisfy~\eqref{optmuepschoice} and $\frac{k_2}{k_1} > \frac{\mu}{\varepsilon}\cdotp$
\end{cor}

\subsubsection{Comparison results}

The positivity of the kernel ensures a first comparison principle.
\begin{prop}\label{prop:univbound}
For $d\geq1$ and any pair of smooth positive functions $f$ and $g$:
\begin{equation}
f\leq g \text{ on }\R_+  \qquad\Longrightarrow\qquad \mathbb{WK}_d[f] \leq  \mathbb{WK}_d[g] \text{ on }\R_+.
\end{equation}
In particular, if $f$ is a quasi-power-law with exponent $-\alpha$ on $[k_1,k_2]$ with a tolerance $\varepsilon_0$, then
there exists $c_0>0$ (given by proposition~\ref{prop:powerlawconsequence}) such that
\begin{equation}
\forall\lambda>0,\qquad
\mathbb{WK}_d[f](\lambda) \geq c_0 e^{-\varepsilon_0} \cdot \mathbb{WK}_d[f] \left[ k^{-\alpha} \mathbf{1}_{[k_1,k_2]} \right](\lambda).
\end{equation}
\end{prop}
Up to a normalization factor, the function $\mathbb{WK}_d[k^{-\alpha} \mathbf{1}_{[k_1,k_2]}]$ is thus
a universal lower bound on $\R_+$ of the Wiener-Khinchin transform of all quasi-homogeneous functions. 
The overall shape of its graph depends on how $\alpha$ compares to 1 and 3 and is depicted
in  figures~\ref{fig:compactsupport}, \ref{fig:smallalpha} and~\ref{fig:largealpha} of~\S\ref{par:numerics}.

\bigskip
The second comparison principle is based on inequalities~\eqref{sandwich}-\eqref{sandwich2} in the appendix.
The graph of $\mathbb{WK}_2[f]$ in log-log coordinates is sandwiched between two copies of that of $\mathbb{WK}_3[f]$
that are only offset vertically by $\log 2$. So even though the previous theorems require some restrictions when $d=2$,
one can use the following comparison principle and the results for $d=3$ to somewhat circumvent those limitations.

\begin{prop}
For any smooth positive function $f$, one has, pointwise:
\begin{equation}
\frac{3}{4} \mathbb{WK}_3[f] \leq \mathbb{WK}_2[f]  \leq \frac{3}{2} \mathbb{WK}_3[f]
\end{equation}
and more generally for $d\geq2$:
\begin{equation}
\frac{d+1}{d+2} \, \mathbb{WK}_{d+1}[f] \leq \mathbb{WK}_d[f]  \leq \frac{d+1}{d} \, \mathbb{WK}_{d+1}[f].
\end{equation}
In log-log coordinates, the vertical offset between the two envelopes is $\log(1+\frac{2}{d})$.
\end{prop}

\subsubsection{Numerical examples}\label{par:numericsexplained}

In section~\ref{par:numerics}, we will investigate the Wiener-Khinchin transform numerically, both to illustrate the previous results
but also to ascertain their sharpness and explore further possible developments.

\paragraph{Transform of a two-regime function.}

For $\alpha>-1$ and $\beta>3$, the function
\begin{equation} \label{tworegimefnct}
f^{\alpha,\beta}_{k_0}(k)=\frac{k^\alpha}{k_0^{\alpha+\beta}+k^{\alpha+\beta}}
\end{equation}
satisfies $\int_0^\infty (1+k^2)f^{\alpha,\beta}_{k_0}(k)dk<\infty$.
This function is clearly asymptotic to $k^\alpha$ at the origin and to $k^{-\beta}$ at infinity,
with a transition around $k=k_0$. One can easily check that
\[
\frac{(K^\sharp_\delta)^{-(\beta-3)}}{\beta-1} \simeq \left( \frac{k_0^{\alpha+3}}{\alpha+3} + \frac{k_0^{-(\beta-3)}}{\beta-3} \right) \delta^{4}
\quad\text{and}\quad
\frac{(K_\eta^\flat)^{\alpha+1}}{\alpha+1} \simeq \left(\frac{k_0^{\alpha+1}}{\alpha+1} + \frac{k_0^{-(\beta-1)}}{\beta-1} \right) \eta^{-(d-1)/2}.
\]
As the thresholds $\delta_0$ and $\eta_0$ depend only on the dimension,
one expects the two asymptotic regimes of $\mathbb{WK}_d[f_{\alpha,\beta;k_0}]$ to ``meet''
around $\lambda\sim k_0^{-1}$.
This behavior is confirmed numerically and illustrated in figure~\ref{fig:tworegimefnct} (p.~\pageref{fig:tworegimefnct}).
In particular, the general profile of $\mathbb{WK}_d[f^{\alpha,\beta}_{k_0}](\lambda)$ does not depend specifically on the values~$\alpha,\beta$;
its only striking feature is the transition from a quadratic growth to a roughly constant state when $\lambda$ crosses $k_0^{-1}$.

\bigskip
When $\beta=3$, the function $f^{\alpha,3}_{k_0}$ remains integrable but fails to have a finite moment of order~2.
The Wiener-Khinchin transform can still be defined as a semi-convergent oscillatory integral. However, the universal quadratic
regime at the origin does not occur anymore. Numerically, one observes a quadratic asymptotic direction in log-log
coordinates (figure~\ref{fig:tworegimefnctlim}) but no asymptote at the origin. This example indicates that the assumptions of theorem~\ref{thm2}
are sharp.

\paragraph{Transform of a three-regime function.}
Let us now consider the following function:
\begin{equation}\label{threeregimefnct}
f(k)=\frac{k^2 \exp(-10^{-4}k)}{(1+k^3)\ln^2(2+k)}\cdotp
\end{equation}
The function $f$ is a custom-built archetype of a ``three regimes'' function:
it behaves like $k^2$ at the origin and until $k\sim 1$;
it has an exponential cut-off after $k\sim 10^4$; in between, it displays a strikingly good quasi-power-law behavior
that matches $k^{-\alpha}$ over almost 4~decades with $\alpha\in(1.5,1.7)$. 
This function plays a central role in~\cite{FV:Num1}.
It is an explicit example of what a realistic energy spectrum of a smooth 3D~solution of Navier-Stokes looks like (see~\S\ref{par:PMotivation})
and is surprisingly close to the best known estimates in the analytic class~\cite{DT1995}, \cite{FV:K41}.

\medskip
Its Wiener-Khinchin transform can be computed numerically and is shown in figure~\ref{fig:threeregimefnct}.
According to theorems~\ref{thm2} and~\ref{thm3}, the transform
admits two universal asymptotic regimes at the origin and at infinity.
In between, as predicted by theorem~\ref{thm4}, it is a quasi-power-law that matches the power-law~$\lambda^{\alpha-1}$ on more than 3~decades.
From the figure~\ref{fig:threeregimefnct}, is also obvious that the introduction of the parameters $\varepsilon$, $\mu$, $\delta$ and $\eta$
was necessary to avoid the two transition regions.

\paragraph{Universal lower bound.}
Proposition~\ref{prop:univbound} depicts $\mathbb{WK}_d[k^{-\alpha}\mathbf{1}_{[k_1,k_2]}]$ as a universal lower bound
of the Wiener-Khinchin transform of any quasi-power-law of exponent $-\alpha$.
Its graph is given in  figure~\ref{fig:compactsupport}. The overall shape is similar to the one in figure~\ref{fig:threeregimefnct},
with one notable difference: as the initial function is supported \textit{away} from zero, its transform presents damped oscillations as $\lambda\to\infty$,
which are kindred to the ones of the cardinal sine function $\frac{\sin \pi \xi}{\pi \xi} = \int_{-1/2}^{1/2} e^{2i\pi x\cdot\xi} dx$.

\medskip
Let us also point out that the theorems \ref{thm2}, \ref{thm3} and \ref{thm4} take a particularly simple form in this case:
\begin{equation}\begin{aligned}
\|\mathbb{WK}_d[k^{-\alpha}\mathbf{1}_{[k_1,k_2]}\|_{\power_1^{2}(0,\delta/k_2)} &\leq A\delta^2\\
\|\mathbb{WK}_d[k^{-\alpha}\mathbf{1}_{[k_1,k_2]}\|_{\power_\infty^{0}(\eta/k_1,\infty)} &\leq B\eta^{-\frac{d-1}{2}}\\
\|\mathbb{WK}_d[k^{-\alpha}\mathbf{1}_{[k_1,k_2]}\|_{\power_\infty^{\alpha-1}(\mu/k_2,\varepsilon/k_1)} &\leq C \varepsilon^{3-\alpha} + D \mu^{-(\alpha-1)}
\end{aligned}\end{equation}
with constants $A,B,C,D$ that only depend on $d$, $\alpha\in(1,3)$, $k_1\simeq K^\flat_\eta$ and $k_2\simeq K^\sharp_\delta$.

\paragraph{About the restriction $1<\alpha<3$ in theorem~\ref{thm4}.}

The range $1<\alpha<3$ was natural in theorem~\ref{thm1} because it conveyed the integrability necessary to define the Wiener-Khinchin transform
of a global quasi-power-law on $(0,\infty)$.
However, if we restrict ourselves to compactly supported quasi-power-laws
(or to quasi-power-laws that have rapidly decaying log-log tails both at zero and at infinity),
integrability is not a problem anymore.

\medskip
In figures~\ref{fig:smallalpha} and \ref{fig:largealpha}, we explore the Wiener-Khinchin transform of $k^{-\alpha}\mathbf{1}_{[k_1,k_2]}$
when $\alpha$ lies outside $(1,3)$. Numerically, it is quite clear that in this case, the overall shape of the graph
of $\mathbb{WK}_d[k^{-\alpha}\mathbf{1}_{[k_1,k_2]}]$  displays only the two universal regimes.
When $\alpha\leq1$, the asymptotically constant regime extends all the way down to $\lambda\sim 1/K^\sharp$.
Note however that  the power-law gauge measuring the flatness on $[1/K^\sharp,1/K^\flat]$ is much larger than the one on $[1/K^\flat,\infty)$.
On the contrary, when $\alpha\geq 3$, the quadratic regime expands from the origin all the way up to $\lambda\sim 1/K^\flat$,
with no noticeable increase of the gauge in the intermediary range.
This numerical evidence suggests that the limitation $1<\alpha<3$ in theorem~\ref{thm4} was necessary.

\paragraph{Transform of multi-regime functions.}
Both for the sake of mathematical exploration and for the possible applications to 2D~turbulence, we also numerically investigated
the Wiener-Khinchin transform of functions that behave as more than one power-law. If one discards the less relevant tails at zero
and at infinity, a typical example is:
\begin{equation}\label{multiregimefnct}
f^{\alpha,\beta}_{k_1,k_2,k_3}(k)=
k^{-\alpha} \mathbf{1}_{[k_1,k_2)} + k_2^{\beta-\alpha} k^{-\beta} \mathbf{1}_{[k_2,k_3]}. 
\end{equation}
We restricted ourselves to the range $\alpha,\beta\in(1,3)$. Numerically, one observes that $\mathbb{WK}_d[f^{\alpha,\beta}_{k_1,k_2,k_3}]$
also presents two quasi-homogeneous ranges, of exponents $\beta-1$ and $\alpha-1$, that happen roughly respectively on the intervals~$[1/k_3,1/k_2]$ and~$[1/k_2,1/k_1]$.
These results are shown on figures~\ref{fig:fourregimefnct1} and~\ref{fig:fourregimefnct2}.
Note that figure~\ref{fig:fourregimefnct2} provides an example of a Wiener-Khinchin transform that is non-concave within the non-constant range $[0,1/K^\flat]$.

\section{Main applications}

The results stated above are of general interest as fine properties of the Fourier transform. However, their main motivation is to clarify
one aspect of the mathematical foundations of hydrodynamic turbulence.

\subsection{Connection with the Fourier transform of radial functions on $\R^d$}\label{par:MMotivation}

The Wiener-Khinchin transform \eqref{main} is closely related to the Fourier transform of radial functions in~$\R^d$.
Let us denote by $|\cdot|$ the Euclidian norm on $\R^d$.
Given a profile $U:(0,+\infty)\to\R$, it is well known  that the Fourier transform on $\R^d$ of the radial function 
$U(|x|)$ is also radial and real-valued:
\begin{equation}\label{eq:radialfourier}
\forall\xi\in\R^d, \qquad  \hat{U}(|\xi|) :=
\int_{\R^d}U(|x|) e^{-2i\pi x\cdot \xi}  dx =2\pi |\xi|^{1-\frac{d}{2}}\int_0^\infty U(\lambda)\lambda^{d/2}J_{\frac{d}{2}-1}(2\pi \lambda |\xi|)d\lambda.
\end{equation}
The function $ k^{\frac{d}{2}-1}\hat{U}(k)$ is the Hankel transform of $\lambda^{\frac{d}{2}-1}U(\lambda)$ and the formula is (formally) self-inverting~:
\begin{equation}\label{eq:inverseradialfourier}
U(\lambda)=2\pi \lambda^{1-\frac{d}{2}}\int_0^\infty \hat{U}(k)k^{d/2}J_{\frac{d}{2}-1}(2\pi \lambda k)dk.
\end{equation}
The connection between the Wiener-Khinchin and the Hankel transforms is given by~\eqref{HankelConnexion}.

\medskip\paragraph{Remark.}
If $k^{d-1}\hat{U}(k)\in L^1(0,+\infty)$ then the inverse Fourier transform can
be computed with the usual integral on $\R^d$;  the inversion formula \eqref{eq:inverseradialfourier} thus holds point-wise.
It is also common knowledge that if $U$ is piecewise continuous and of bounded variation in every finite subinterval of $(0,+\infty)$ then the inversion
formula holds at each point of continuity of $U$.
See~\cite[p.155]{Stein} and \cite{Katz} for further details on the Fourier transform of radial functions, spherical harmonics and questions of convergence.

\subsection{Physical motivation: applications to fluid mechanics and turbulence}\label{par:PMotivation}

In physics, the Wiener-Khinchin transform occurs naturally to describe the relation between the \textit{energy spectrum} and the \textit{structure function}
of a signal \cite{W30}, \cite{Kn34}, \cite{W49}.
Let us first recall briefly, in a general setting, the definition of these fundamental notions of mathematical physics.

\subsubsection{Correlation, structure function and energy spectrum of a signal}

Let us consider a square integrable function $u:\Omega \to \R^q$.
In the subsequent applications, one will have either $q=1$ ($u$ is a scalar function) or $q=d$ ($u$ is a vector field on $\R^d$).
To avoid  the spectral subtleties of a general domain $\Omega$, one will assume that $\Omega=\R^d$. For precise definitions on $\Omega=\T^d$, see~\eg\cite{FV:K41}.
In what follows, $|z|=(\sum z_j^2)^{1/2}$ denotes the Euclidian norm of $z\in\R^n$ for any integer $n$.

\begin{dfn}
The \textbf{correlation function} $R:\R^d\to\R$ of $u$ is the trace of the correlation matrix:
\begin{equation}
\forall y\in\R^d, \qquad
R(y)=\sum_{j=1}^q  \int_{\R^d} u_j(x)u_j(x+y)dx.
\end{equation}
\end{dfn}
The Fourier transform of the correlation function is exceptionally simple:
\begin{equation}
\hat R(\eta) := \int_{\R^d} R(y) e^{-2i\pi y\cdot\eta} dy = |\hat{u}(\eta)|^2.
\end{equation}
In particular, as the correlation function $R$ can be computed as the inverse Fourier transform of $|\hat{u}|^2$, it
does not contain any information on the $2q-1$ phases within~$\hat{u}:\R^d\to\C^q$.
With the applications in mind, let us also observe the two following facts.
\begin{enumerate}
\item
A typical assumption on a velocity field of hydrodynamics is $u\in L^2$ (flow of finite energy). It implies that $R\in \mathcal{F}(L^1)$.
In this case, the correlation function is uniformly continuous, bounded pointwise by $\|u\|_{L^2}^2$ and tends to zero at infinity.
\item
On the other hand, if $u\in L^1$ (flow of finite momentum), then $R\in L^1$ too and $\|R\|_{L^1} \leq \|u\|_{L^1}^2$.
\end{enumerate}

\begin{figure}[H]
\captionsetup{width=.7\linewidth}
\begin{center}
\includegraphics[width=.7\linewidth]{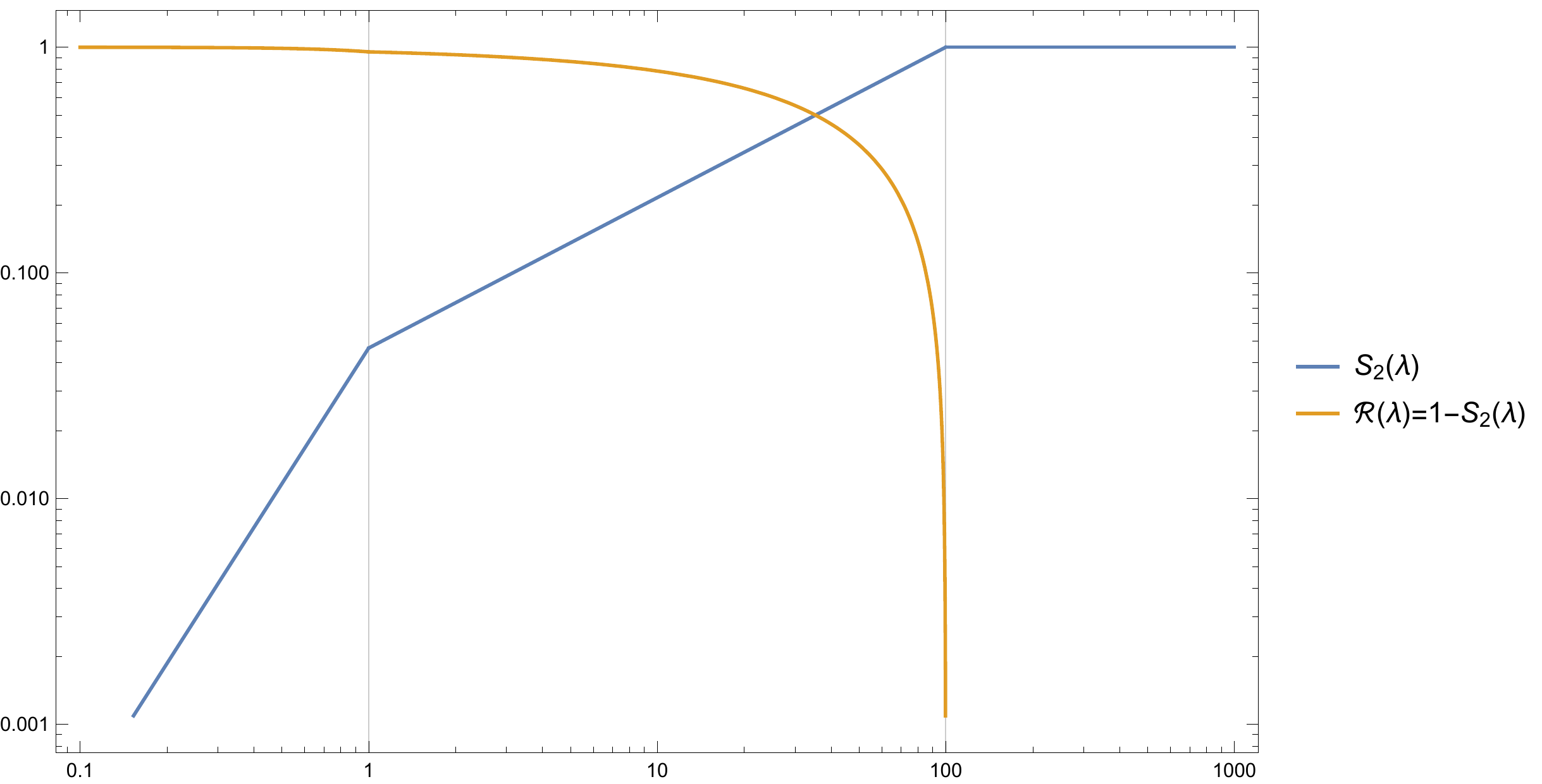}
\caption{\label{fig:scalarCorrelation}\small
A mockup of a typical structure function $S_2(\lambda)$ and the corresponding scalar correlation $\mathcal{R}(\lambda)$. 
Note the radically different profiles.
}\end{center}
\end{figure}

\begin{figure}[H]
\captionsetup{width=.85\linewidth}
\begin{center}
\includegraphics[width=.85\linewidth]{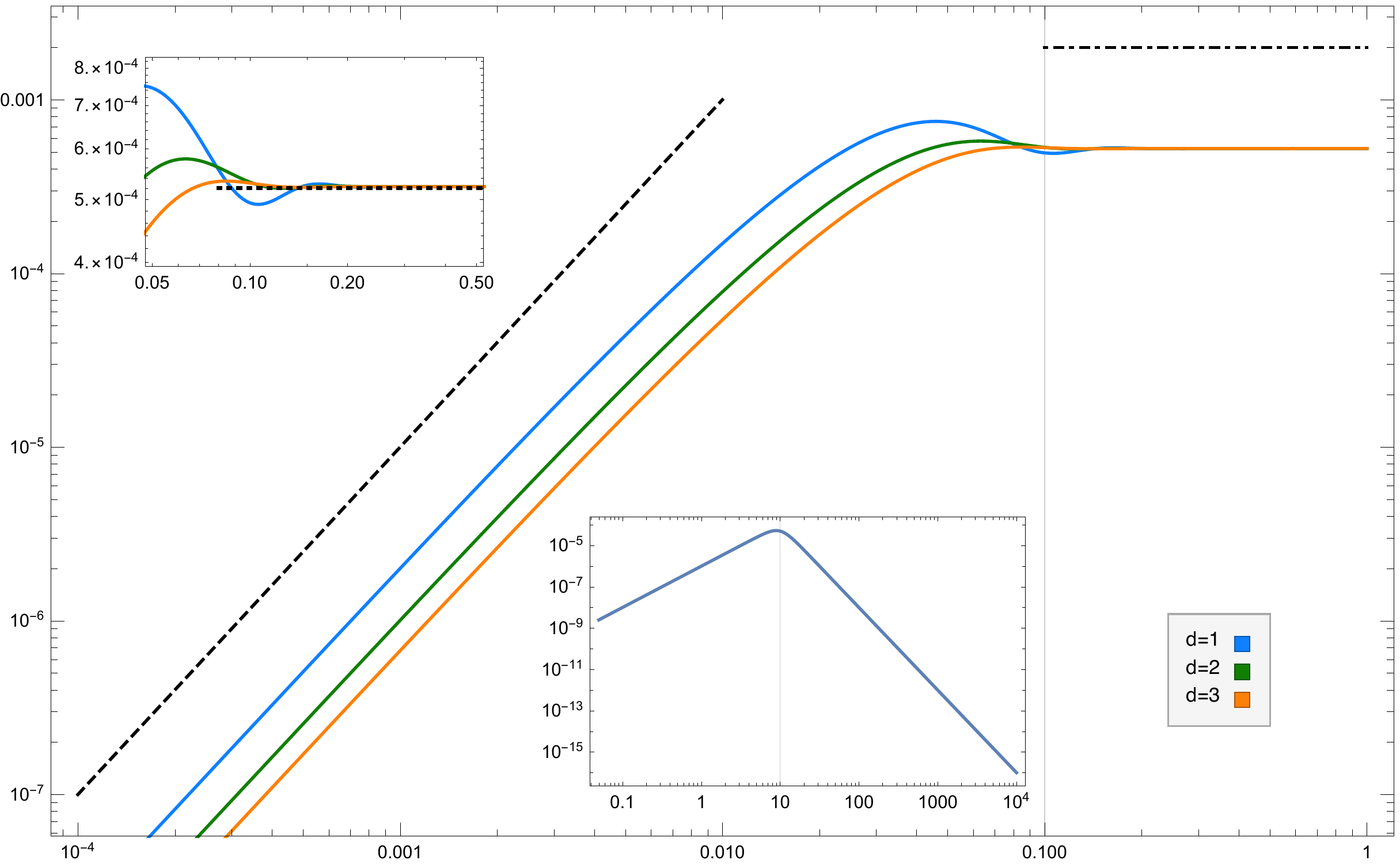}
\caption{\label{fig:tworegimefnct}\small
Wiener-Khinchin transforms of a two-regime function $f^{2,4}_{10}$ defined by~\eqref{tworegimefnct}, in dimensions $d=1,2,3$.
The reference slopes 2 and 0 are given by the dashed lines.
The graph of~$f$ is in the bottom inset.
The vertical delimiter marks the value $k_0$ at which~$f$ reaches its maximum, and the corresponding $k_0^{-1}$ threshold.
The top-left inset is a zoom on $\mathbb{WK}_d[f](\lambda)$ for large $\lambda$.}
\end{center}
\end{figure}

\begin{figure}[H]
\captionsetup{width=.85\linewidth}
\begin{center}
\includegraphics[width=.85\linewidth]{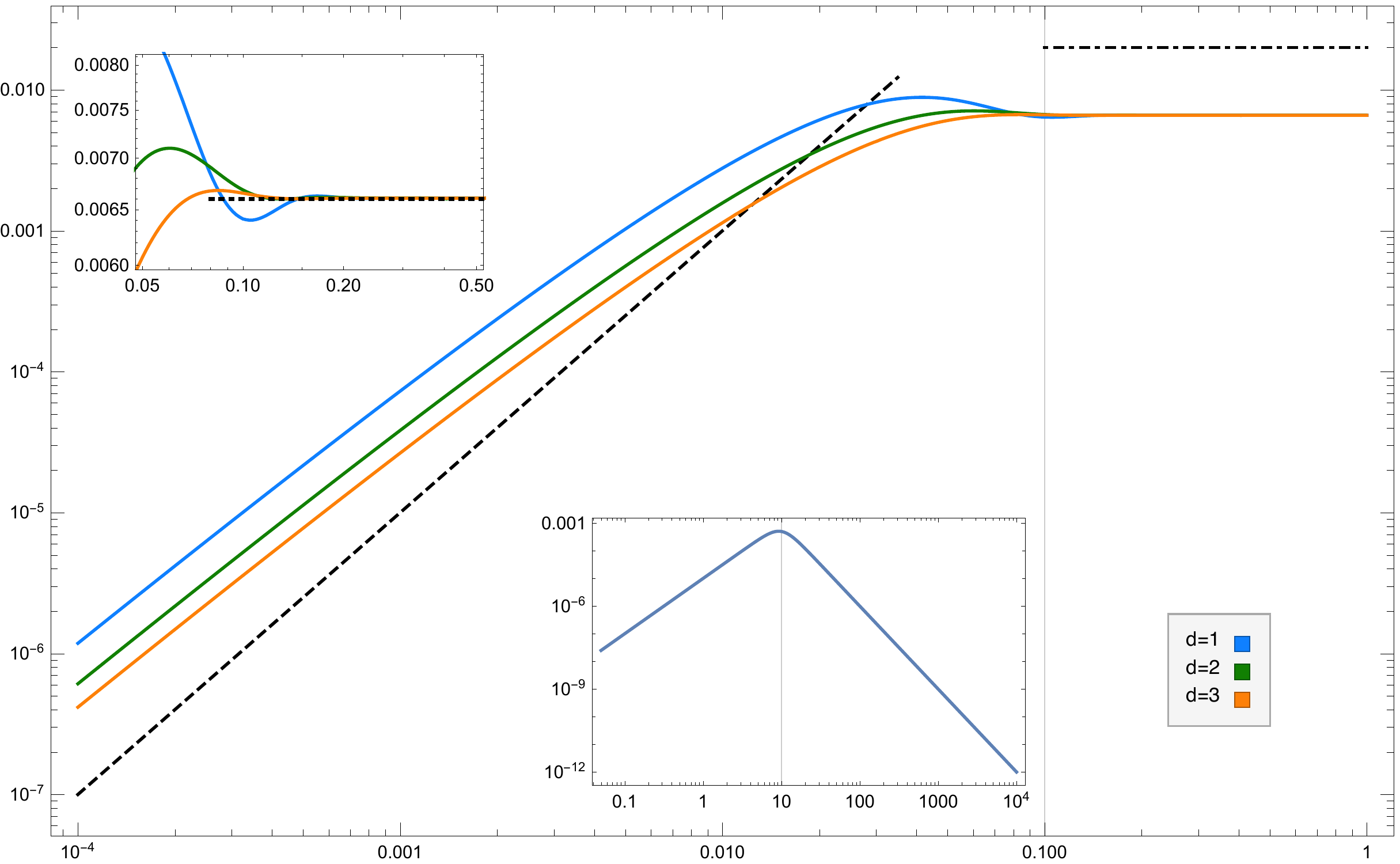}
\caption{\label{fig:tworegimefnctlim}\small
Wiener-Khinchin transforms of \eqref{tworegimefnct} in a limiting case.
The layout is the same as in figure~\ref{fig:tworegimefnct} and the reference slopes are still 2 and 0.
The fact that $f^{2,3}_{10}$ does not have a finite second order momentum prevents it from having a quadratic
asymptote at the origin.}
\end{center}
\end{figure}

\begin{dfn}
The \textbf{second (or $L^2$-based) structure function} is the radial average of the $L^2$-increments of $u$. It is defined for $\lambda\in(0,+\infty)$ by:
\begin{equation}\label{S2incr}
S_2(\lambda)=\frac{1}{2}\int_{\mathbb{S}^{d-1}}\int_{\R^d} |u(x+\lambda\vartheta)-u(x)|^2  \frac{dx \, d\vartheta}{|\mathbb{S}^{d-1}|}\cdotp
\end{equation}
\end{dfn}
If $u\in L^2$, the second structure function relates to the radial averages of the correlation function, which is sometimes called
the \textbf{scalar corelation function} (see figure~\ref{fig:scalarCorrelation}):
\begin{equation}\label{S2Gamma}
\forall \lambda>0,\qquad
\mathcal{R}(\lambda) := \frac{1}{|\mathbb{S}^{d-1}|}\int_{\mathbb{S}^{d-1}} R(\lambda\vartheta)  d\vartheta = \|u\|_{L^2}^2 - S_2(\lambda).
\end{equation}
Thanks to the first observation just above, $S_2$ is positive, uniformly continuous  and $\lim\limits_{\lambda\to+\infty}S_2(\lambda)=\|u\|_{L^2}^2$.
If $u\in L^1\cap L^2$, then, using~\eqref{S2Gamma} and the second observation, the rate of convergence is controlled by:
\begin{equation}\label{univConstFluids}
|\mathbb{S}^{d-1}|\, \int_0^\infty \left| \|u\|_{L^2}^2-S_2(\lambda)\right|  \lambda^{d-1}d\lambda = \|R\|_{L^1} \leq \|u\|_{L^1}^2.
\end{equation}
In anticipation of~\eqref{eq:WK} below, let us point out that \eqref{univConstFluids} provides a hint of the convergence towards the universal constant state
at infinity that one should expect from any Wiener-Khinchin transform.

\begin{dfn}
The \textbf{energy spectrum} of $u$ is the radial total of $|\hat{u}|^2$. It is defined on $\R_+$ by:
\begin{equation}\label{energyspectrum}
E(k) = \int_{k\mathbb{S}^{d-1}} \!\! |\hat{u}|^2  
= k^{d-1}\int_{\mathbb{S}^{d-1}} \hat R(k\vartheta) d\vartheta.
\end{equation}
Note that $E(k)$ depends on the implicit variables of $u$, and in particular on $t$ (unless $u$ is a time average).
\end{dfn}
For $u\in L^2(\R^d)$, the Bessel identity means that the energy spectrum is integrable and that
\begin{equation}\label{bessel}
\|u\|_{L^2}^2 = \int_0^\infty E(k) dk.
\end{equation}
In general, the energy spectrum has no reason to have a better regularity than $L^1(0,+\infty)$.
However, if  $u\in L^1\cap L^2$, then $\hat{u}\in \mathcal{F}(L^1)$ and $E$ becomes continuous, bounded pointwise by 
$|\mathbb{S}^{d-1}|  \|u\|_{L^1}^2 K^{d-1}$.
In dimension $d=3$, one gets $E(K) \lesssim K^2$ near the origin and when this upper bound is sharp, it is called a \textit{Saffman spectrum}.

\paragraph{Example.} \label{ex:localisationofflow}
For Leray solutions of Navier-Stokes stemming from $u_0\in L^1\cap L^2$, one gets~\cite{profile:BFV}, \cite{FV:K41}:
\[
E(K,t) \lesssim  
\inf_{t'<t}\left( \|u(t')\|_{L^1}^2+C\nu^{-1}(t-t') \|u(t')\|_{L^2}^4 \right) K^{d-1}.
\]
Provided $u_0$ is well localized, the following asymptotic profile is also known (again, see \cite{profile:BFV}):
\[
u(t,x) =e^{t\Delta}u_0(x)+ \gamma_d \nabla\left(\sum_{i,j}\frac{\delta_{i,j}|x|^2-dx_ix_j}{|x|^{d+2}}\int_0^t\int_{\R^d} u_i(\tau,\eta)u_j(\tau,\eta)d\tau d\eta\right)+\mathfrak{o}\left(\frac{1}{|x|^{d+1}}\right).
\]
The decay $|u|\lesssim \|u\|_{L^2}^2/|x|^{d+1}$ implies $\hat{u} \in C^{0,s}(\R^d)$ for any $0< s<1$.
The energy spectrum $E(K)$ is then also in the H\"older class $C^{0,s}(\R_+)$.
It has been shown in \cite{FV:K41} that when this asymptotic profile holds, it provides tighter upper bounds on the lower-end of the energy spectrum, namely:
\[
\forall \beta<d+1, \qquad E(K)\leq C_\beta(t) K^{\beta}.
\]
\textit{Batchelor's spectrum} (\ie $\beta=d+1=4$ in dimension $3$) is a generically inaccessible endpoint because the profile requires a slightly stronger condition
on the flow \cite{BM02} \eg$|x|u(t,x)\in L^1(\R^3)$, which turns out to require the energy matrix $(u_i\vert u_j)$ to remain constant and equal to $\frac{1}{3}I_3$, which is generically not true as explained in~\cite{B04a}, \cite{B04b}, \cite{profile:BFV}. However, when the flow does have the necessary symmetries,
the previous estimate of $E(K)$ still holds for $\beta=d+1$, but with an additional $|\log(K/K_0)|$ factor.
\cqfd

\bigskip
The main motivation for studying the Wiener-Khinchin transform is the following.
\begin{thm}[Wiener-Khinchin \cite{W30}, \cite{Kn34}]
For any $u\in L^2(\R^d;\R^q)$, one has:
\begin{equation}\label{eq:WK}
S_2 =  \mathbb{WK}_d[E].
\end{equation}
\end{thm}
This result is part of folklore and usually stated in the context of the analysis of  time-series,
but a deterministic and self-contained proof is short enough to be included here.
\Proof
The key is to apply the inversion formula \eqref{eq:inverseradialfourier} to the pair of radial Fourier profiles
\[
U(\lambda)=\int_{\mathbb{S}^{d-1}} R(\lambda\vartheta) d\vartheta = |\mathbb{S}^{d-1}|\mathcal{R}(\lambda)
\qquad\text{and}\qquad
\hat{U}(k)=\int_{\mathbb{S}^{d-1}} \hat{R}(k\vartheta) d\vartheta=\frac{E(k)}{k^{d-1}}
\cdotp
\]
According to the remarks that follow \eqref{eq:inverseradialfourier}, the inversion formula is valid point-wise under the sole assumption that $u\in L^2(\R^d)$ because then
$k^{d-1}\hat{U}(k) = E(k)\in L^1(0,\infty)$. One gets:
\[
 |\mathbb{S}^{d-1}|\mathcal{R}(\lambda) 
 =2\pi \lambda^{1-\frac{d}{2}}\int_0^\infty E(k) k^{1-\frac{d}{2}}J_{\frac{d}{2}-1}(2\pi \lambda k)dk.
 \]
Combined with the identities \eqref{S2Gamma} and \eqref{bessel}, the computation boils down to a deterministic version of the celebrated Wiener-Khinchin formula:
\begin{equation}
S_2(\lambda) = \int_0^\infty \left(1-\frac{2\pi}{|\mathbb{S}^{d-1}|}(\lambda k)^{1-\frac{d}{2}}J_{\frac{d}{2}-1}(2\pi \lambda k)\right) E(k) dk.
\end{equation}
As the area of the unit sphere is $|\mathbb{S}^{d-1}| = 2 \pi^{d/2} / \Gamma\left(\tfrac{d}{2}\right)$, the result is established.
\cqfd

\paragraph{Remark.}
Formula \eqref{eq:radialfourier} provides the converse identity, which is often used as a practical definition of the energy spectrum
(note that~\eqref{univConstFluids} ensures the convergence if $u\in L^1\cap L^2$):
\begin{equation}\label{eq:reconstruction}
E(K) = \frac{4\pi}{\Gamma\left(d/2
\right)} \int_0^\infty  (\pi\lambda K)^{d/2} J_{\frac{d}{2}-1}(2\pi \lambda K) \mathcal{R}(\lambda) d\lambda
\end{equation}
with $\mathcal{R}(\lambda) = \|u\|_{L^2}^2-S_2(\lambda) $. Figure~\ref{fig:scalarCorrelation} illustrates 
the important fact that contrary to $S_2(\lambda)$, one should not expect $\mathcal{R}(\lambda)$ to be a quasi-power-law,
except for the quasi-constant regime near the origin.

\subsubsection{Homogeneous turbulence, in the sense of Kolmogorov, in a nutshell}

Turbulence is a collection of qualitative and quantitative properties that are commonly observed at the intermediary scales of generic, highly agitated, viscous flows.
A practical criterion to detect \textit{homogeneous turbulence, in the sense of Kolmogorov,} on a time interval $[T_0,T_1]$ 
is the existence of a so-called \textit{inertial range}: it is either a large spectral interval $[k_1,k_2]$ along which the time average of the energy spectrum~\eqref{energyspectrum}
is a quasi-power-law in the sense of definition~\ref{HFR}, or a large spatial interval $[\lambda_1,\lambda_2]$ along which the time~average of the second
structure function~\eqref{S2incr} is also a quasi-power-law. The respective exponents given by a scaling argument (\ie dimensional analysis) and confirmed
experimentally (\eg in wind tunnels) are $k^{-5/3}$ and $\lambda^{2/3}$ for 3D flows.
The \textit{Reynolds number} is (a power of) the dimensionless ratio $k_2/k_1$ or $\lambda_2/\lambda_1$.

\medskip
For an introductory review of turbulence, I recommend for example H.K.~Moffatt's article \cite{Moff2012} or A.~Tsinober's  book~\cite{Tsin09}.
For the engineering aspects, P.A.~Davidson's book \cite{Davidson04} is accessible and is quite attentive to mathematical correctness.
The spectral point of view of the founding papers of A.N.~Kolmogorov \cite{K41a}, \cite{K41b}, \cite{K41c} and A.~Obukov \cite{Obukov}
is carefully exposed in U.~Frisch's book~\cite{Frisch95} and has been recently revisited by R.~Lewandowsky~\cite{LewanI},  \cite{LewanII}.
L.~Onsager's point of view is complementary to that of Kolmogorov and was reviewed by G.L.~Eyink and K.~Sreenivasan~\cite{ES06}.
We will not address here the more advanced questions of intermittency and the multifractal models, \eg the celebrated log-normal model
of Kolmogorov. Nor will we address  the subject of universal attractors, degrees of freedom, or the question of the closure
of hierarchy models (see \eg the references in~\cite{Frisch95}, \cite{Davidson04}, \cite{Tsin09}).

\medskip
At a finite Reynolds number, Kolmogorov's theory is compatible with an analytic regularity of the flow because
most of the energy/enstrophy (\ie Sobolev norms) is concentrated on a \textit{finite} number of scales.
However, as $\Re\textrm{e}\to\infty$, the inertial range grows indefinitely (\ie $k_2\to\infty$ or $\lambda_1\to 0$), which
suggests that fully developed turbulence might be the echo of singularities of the Euler equation.
Onsager went one step further and conjectured a relation between singularity and non-viscous dissipation. In naive terms,
it says that if the flow is not regular enough, then the usual energy balance might be spontaneously violated (see \eg J.~Duchon, R.~Robert~\cite{DR00}).
After about 70 years and numerous contributions by prominent mathematicians, this conjecture was finally solved 
by P.~Isett~\cite{I2018}.
Since then, the associated convex integration scheme inspired T.~Buckmaster and V.~Vicol \cite{BV19},
along with M.~Colombo~\cite{BCV19},
to build wild viscous solutions of Navier-Stokes. The current big picture is the following: 
an extremely large family of weak (in the distributional sense), non-unique solutions, with singular dissipation, swarms
around each Leray solution, and in particular in the neighborhood of each smooth solution,
even though these remain unique in the Leray class.

It is remarkable that the building blocks of this construction involve ideas rooted in the physics of turbulence.
However, the scales at which they are put into play seem to be, for now, far beyond the realm of validity of real-life
hydrodynamics\footnote{The exponential-tower growth of the frequencies of each corrector term in \cite{BV19}-\cite{BCV19} would undoubtedly reach scales
that are far below the magnitude of the mean free path between fluid molecules, which is only a few Angstr\"oms, \ie $10^{-10}\textrm{m}$.}.

\medskip
To put this mathematical breakthrough in perspective, let us ask the following retorical question:
\begin{quote}
\textit{If one pumps enough energy into a viscous fluid to generate turbulence before
letting it go back to rest, where has the energy gone?
Was it just dissipated into heat by molecular friction or has some of it been ``robbed'' by a low-scale mathematical trick unrelated to viscosity?}
\end{quote}
The former is the only physically admissible possibility, but it requires the flow to be, at least, a Leray solution
and all of the energy to be accounted for.
The latter is only possible if the flow happens to be a wild solution of Navier-Stokes (which is not a fantasy anymore).
But this option violates the most fundamental law of physics and will likely be rejected as a failure of the Navier-Stokes model itself as an accurate description of physics.
To ``save'' Navier-Stokes as a physically relevant model, it is therefore urgent that we develop a theory of turbulence at high but finite Reynolds numbers
that is coherent within the restricted realm of Leray solutions.

\medskip
Many authors in physics and in mathematical physics
assert that one  should not over-emphasize the spectral representation of turbulence or they simply do not use it,
\eg\cite{CCALPR2012}, \cite{Wall2014},
\cite{dSMWM2015}, \cite{Ch2019}.
Part of the picture is the fact that,  in the experiments,
the spatial increments can be measured directly with hot-wire anemometer techniques.
On the contrary, energy spectra are usually only indirectly accessible
via the scalar correlation function~\eqref{S2Gamma} and~\eqref{eq:reconstruction},
or by applying a windowed Fourier transform to cut away the upstream subregion that produced the turbulence,
or through wavelet techniques (see \eg\cite{DSKFSDD2018}).

A deeper objection to the spectral representation is the fact that it is often used to express an oversimplified version of the ``cascade'' idea.
Even if our everyday-life experience confirms that 3D turbulent eddies break down from 
large ones to small ones\footnote{L.~Richardson's poem in 1922: \textit{Big whorls have little whorls,
which feed on their velocity. And little whorls have lesser whorls, and so on to viscosity.}},
turbulence is not just a stepwise cartoon and the phase-space nature of it is somewhat more subtle (see \eg\cite{JLA2007}, \cite{KKSK01}, \cite{FARGE07} 
or  \cite{CONSTANTIN94} and the literature on depleted coherent structures). At a given frequency (\ie eddy size), a realistic turbulent flow
is not necessarily statistically invariant under very small translations or rotations (think of the gap between two vortex filaments).
In other words, the cascade is not just a matter of Fourier modes, but is, in essence, micro-local.


\subsubsection{Two practical definitions of turbulence}

In this section, and to keep things simple,
$u:[0,T]\times \R^3 \to \R^3$ denotes a Leray solution to the incompressible Navier-Stokes equation in $\R^d$ for a fluid of constant density:
\begin{equation}\label{NS}
\begin{cases}
\partial_t u - \nu\Delta u + u\cdot\nabla u = \nabla p\\
\operatorname{div} u = 0
\end{cases}
\end{equation}
where $\nu>0$ is the kinematic viscosity and $p:[0,T]\times \R^3 \to \R$ is the pressure field.
Let us also introduce two fixed dimensionless parameters $\Re_0\gg1$ and $\gamma_0\ll1$.

\bigskip
The first practical definition of Kolmogorov's homogeneous turbulence is the following spectral one.
\begin{dfn}[see \cite{FV:K41}]\label{def:turbSpectral}
A velocity field $u:[T_0,T_1]\times\R^3\to\R^3$ 
with finite energy and momentum (i.e. $u\in L^1\cap L^2$)
is called \textbf{homogeneously turbulent in the
spectral sense} (or \textbf{spectrally turbulent} for short) if there exists $k_1<k_2$  such that the following conditions are satisfied.
\begin{enumerate}
\begin{subequations}
\item The average energy spectrum $\textstyle \bar{E}(k)=\frac{1}{T_1-T_0}\int_{T_0}^{T_1} E(t,k)dt $
is a quasi-power-law of exponent~$-5/3$ on~$[k_1,k_2]$, up to the tolerance $\gamma_0$, i.e.:
\begin{equation}\label{K41small}
\|\bar{E}\|_{\power_0^{-5/3}(k_1,k_2)} \leq \gamma_0.
\end{equation}
\item The spectral ``inertial range'' $[k_1,k_2]$ is large:
\begin{equation}\label{K41large}
\frac{k_2}{k_1}\geq \Re_0.
\end{equation}
\item A somewhat substantial part of the dissipation occurs in the inertial range:
\begin{equation}\label{K41split}
\int_0^\infty k^2 \bar{E}(k) dk < 3 \int_{k_1}^{k_2} k^2 \bar{E}(k)dk.
\end{equation}
\end{subequations}
\end{enumerate}
\end{dfn}
One can then show~\cite{FV:K41} that the so-called \textit{integral scale} $k_1^{-1}$ necessarily relates to the overall average volume of the largest eddies through the
fundamental (but not well-known) formula:
\begin{equation}
k_1^{-3} \simeq \operatorname{Vol}(u ; T_0,T_1) :=
\frac{\int_{T_0}^{T_1} \|u(t)\|_{L^1}^2 dt}{\int_{T_0}^{T_1} \|u(t)\|_{L^2}^2 dt}\cdotp
\end{equation}
To deal with  real-life domains where  homogeneous turbulence only occurs within a subregion, one could localize the definition~\eqref{def:turbSpectral} to the relevant sub-region of
the flow by replacing $u$ by $\chi u$ where $\chi:\R^3\to[0,1]$ is a smooth cut-off function and then, for example,
use \eqref{S2Gamma} and~\eqref{eq:reconstruction} to define the energy spectrum of the excised flow.

\paragraph{Remarks.}
\begin{enumerate}
\item Let us point out that, for convenience, assumption~\eqref{K41small} was replaced  in \cite{FV:K41} by the slightly stronger one
$\|\bar{E}\|_{\power_\infty^{-5/3}(k_1,k_2)} \log\left(k_2/k_1\right) \leq \gamma_0$,
which also emphasises the subtle balance between the tolerance~$\gamma_0$ and the Reynolds number~$\Re\textrm{e}=(k_2/k_1)^{4/3}$.
In this case, one says that $u$ is \textbf{strongly homogeneously turbulent in the spectral sense}.
\item
In the classical picture of turbulence, \eg\cite{Batch53},
it is customary to expect the energy and dissipation scales to be disjoint (\ie the zones contributing the most
to the $L^2$ and $H^1$ norms are spectrally disjoint).
The numerical value of the constant in the assumption~\eqref{K41split} is ultimately irrelevant, but its finiteness acknowledges the fact that
part of the dissipation will \textit{always} occur in the inertial range. This fact  is not only physically meaningful (as strongly emphasized by~\cite{Frisch95})
but it is also a mathematical necessity that bears control over the tails of the energy spectrum outside of the inertial range.
\end{enumerate}

In \cite{FV:K41}, it was also shown that this definition (and a similar one on $\mathbb{T}^3$) is coherent
and that it allows one to prove  the well-known results that are part of the folklore of turbulence (formulas for $k_1$ and $k_2$ in terms
of the total energy, viscosity and dissipation rate, \ldots).
This definition also implies that \textit{the dissipation rate of spectrally turbulent Leray solutions is asymptotically independent
of the viscosity} as $\Re\textrm{e}=(k_2/k_1)^{4/3}\to\infty$ because of the formula \cite[Thm.~7.1]{FV:K41}:
\begin{equation}\label{asymptDissipation}
\varepsilon:= \nu \int_0^\infty K^2 \bar{E}(K) dK \propto \frac{U^3}{L}
\quad\text{with}\quad
U = \sqrt{\int_0^\infty \bar{E}(K)dK}, \quad L=\operatorname{Vol}(u ; T_0,T_1)^{1/3}
\end{equation}
which only involves the total energy and the integral scale and not (directly) the viscosity.
In particular, definition~\ref{def:turbSpectral} implies an a-priori estimate of the half-life of turbulence: \textit{the time necessary for
a smooth\footnote{When $u$ is smooth, $\varepsilon$ defined by~\eqref{asymptDissipation} coincides with the time average of $-\frac{d}{dt}\|u(t,\cdot)\|_{L^2}^2$
\ie it is \textsl{exactly} the average dissipation rate of energy:
\[
\varepsilon=\frac{ \|u(T_0)\|_{L^2}^2 - \|u(T_1)\|_{L^2}^2 }{ T_1-T_0 }\cdotp
\]}
but spectrally turbulent solution to dissipate exactly half of its kinetic energy} is
\begin{equation}
T_1-T_0 \simeq \frac{L}{U}\cdotp
\end{equation}
and thus coincides with the so-called \textit{large eddy turnover time}.

\medskip
Let us underline, however,  that the question of proving the existence of spectrally turbulent solutions of Navier-Stokes
is  an \textbf{open problem}. At least on $\R^3$ and $\mathbb{T}^3$, such solutions,  if they exist, are expected to saturate
the current best analytic regularity estimates, up to a logarithmic factor (see~\cite{DT1995}, \cite{FV:K41}, \cite{FV:Num1} and the example~\eqref{threeregimefnct} above).

\bigskip
Without further ado, here is an alternate practical definition of turbulence, based on $L^2$-increments.
 \begin{dfn}\label{def:turbSpatial}
A velocity field $u:[T_0,T_1]\times\R^3\to\R^3$  with finite energy (i.e. $u\in L^2$)
is called \textbf{homogeneously turbulent in the sense of $L^2$-increments}
if there exists $\lambda_1<\lambda_2$  such that the following conditions are satisfied.
\begin{enumerate}
\begin{subequations}
\item The structure function \eqref{S2incr} averaged on $[T_0,T_1]$
is a quasi-power-law of exponent~$2/3$ on~$[\lambda_1,\lambda_2]$, up to the tolerance $\gamma_0$, i.e.:
\begin{equation}
\|\bar{S}_2\|_{\power_0^{2/3}(\lambda_1,\lambda_2)} \leq \gamma_0.
\end{equation}
\item The ``inertial range'' $[\lambda_1,\lambda_2]$ is large:
\begin{equation}
\frac{\lambda_2}{\lambda_1}\geq \Re_0.
\end{equation}
\end{subequations}
\end{enumerate}
\end{dfn}
A third condition, similar to~\eqref{K41split}, that expresses a form of scale separation between energy and enstrophy
might be necessary at some point in the future but I don't have a satisfactory formulation for it yet. 

\paragraph{Remark.}
Investigating the $L^p$-based increments is a central task for understanding intermittency.
According to the celebrated $4/5$ law \cite{Frisch95}, one could also seek an $L^3$-based definition where $\bar{S}_3$
(or some variant where the increments are only parallel to the local mainstream) is measured in the gauge $\power_0^1(\lambda_1,\lambda_2)$.
But this question is beyond the scope of the present article.

\subsubsection{Consequences of our results on the mathematical foundations of fluid turbulence}

Even in excellent physics textbooks (\eg\cite{Frisch95}, \cite{Davidson04},\ldots), it is always stated as ``obvious'' that the two
definitions \ref{def:turbSpectral} and \ref{def:turbSpatial} are equivalent.
However, the justification never goes beyond observing that these power-laws form a Fourier pair of conjugate exponents.

\medskip
Theorem~\ref{thm1} quantifies the global stability of the Wiener-Khinchin pairs $(k^{-\alpha},\lambda^{\alpha-1})$ for very small perturbations
and can be seen as a formal version of the heuristic argument of the textbooks.
However, in a realistic energy spectrum, most of the energy is
concentrated on a \textit{large but finite} number of scales. This number is directly related to the number of degrees of freedom of the turbulent system.
This means that the tails of the energy spectrum differ radically from a homogeneous function.

Applying theorem~\ref{thm1} to real-world energy spectra would therefore not give any significant upper bound on the structure function.
In other words, theorem~\ref{thm1} (or the cheap Fourier-pair property)
cannot be used to prove that a spectrally turbulent flow is also turbulent in the sense of $L^2$-increments.

\medskip
Theorem~\ref{thm4} investigates this exact situation and can be applied to realistic energy spectra.
\begin{cor}\label{maincor}
If $u\in L^1\cap L^2$ is a strongly homogeneously turbulent field in the spectral sense on the time interval $[T_0,T_1]$,
with a spectral inertial range $[k_1,k_2]$ and a tolerance $\gamma_0$, and if
\begin{equation}
\frac{k_2}{k_1} > \inf \left\{ \frac{\mu}{\varepsilon} \,;\, \sigma_{5/3}(\varepsilon,\mu) \text{ satisfies }\eqref{optmuepschoice}\right\},
\end{equation}
then $u$ is also a homogeneously turbulent field in the sense of $L^2$-increments, on the
reduced dual inertial range $[\lambda_1,\lambda_2]=[\mu/k_2,\varepsilon/k_1]$ and with a tolerance $2\gamma_0$.
\end{cor}
In naive terms, the potential damages that the tails of the energy spectrum can do to the ideal homogeneous Fourier pair are kept in check: 
\textit{if the inertial range of the energy spectrum is large enough, then the structure function will also display an inertial range at length scales
that are roughly the inverse of the inertial frequencies}, but maybe with a slightly shorter span. Let us point out that
this reduction of the range has been experimentally observed, for example, in~\cite{HSLFGL2010}.

\medskip
Note that we have only adressed here the first half of the equivalence between the definitions~\ref{def:turbSpectral} and~\ref{def:turbSpatial}.
The converse implication would require us to control the Hankel-type transform \eqref{eq:reconstruction} whose kernel is highly oscillatory and
grows as $\lambda^{(d-1)/2}$ so it is not directly accessible with the results exposed here. It will be addressed later.
On the same subject, see also the alternate approach \cite{LewanI}, \cite{LewanII}, which follows more closely Kolmogorov's original assumptions \cite{K41a},
\cite{K41b}, \cite{K41c}.

\bigskip
Let us conclude this section with a brief mention of some other turbulence models for which exponents of the energy spectrum other than $-5/3$ occur.
All exponents of $k^{-\alpha}$ mentioned in table~\ref{FIG2}  belong to the range $1<\alpha<3$ and are thus subject to theorem~\ref{thm4}
and corollary~\ref{maincor}.
One can therefore claim the same qualitative conclusions on the relation between energy spectra and the corresponding structure functions for all of these models.

\begin{table}[H]
\captionsetup{width=.6\linewidth}
\begin{center}\begin{tabular}{lc}\hline
\bf Model & $k \frac{d}{dk} (\log E(k))$\\[1pt]\hline
Kolmogorov  & $-5/3 \simeq -1.67$ \\\hline
Intermittency of Kolmogorov turbulence & -1.7 \\\hline
Driven supersonic MHD turbulence & -1.74\\\hline
Observation in molecular clouds & -1.76\\\hline
Solenoidal forcing of turbulence & -1.86\\\hline
Compressive forcing of turbulence   & -1.94\\\hline
Second observation in molecular clouds  & -1.94\\\hline
Burgers turbulence & -2\\\hline
\end{tabular}
\caption{\label{FIG2}\small
Exponent  of the energy spectrum in the inertial range for various turbulence models \cite{SS13}.}
\end{center}
\end{table}

\section{Global duality of quasi-power-laws (proof of theorem~\ref{thm1})}
\label{proof:thm1}

First, we want to obtain estimates between
the global gauges $\power_i^\theta(0,\infty)$
of $f$ and $\mathbb{WK}_d[f]$ when $d\geq2$.

\subsection{Stability of the log-log slope $\power_\infty$}
By integration by part, $\mathbb{WK}_d[f]$ satisfies the following ODE:
\begin{equation}\label{ODE1}
t \mathbb{WK}_d[f]'(t) + \mathbb{WK}_d[f](t) = - \int_0^\infty H_d(st) sf'(s) ds.
\end{equation}
Let us define $\varepsilon(s)$ by $s f'(s) = (-\alpha + \varepsilon(s)) f(s)$;
the ODE becomes:
$$t \mathbb{WK}_d[f]'(t) -(\alpha-1) \mathbb{WK}_d[f](t) = -\int_0^\infty H_d(st)\varepsilon(s)f(s)ds.$$
As $d\geq2$, the kernel $H_d$ is positive so the right-hand side is bounded by $\|\varepsilon\|_{L^\infty}  \mathbb{WK}_d[f](t)$.
As $\|\varepsilon\|_{L^\infty}$ coincides with $\|f\|_{\power^{-\alpha}_\infty(0,\infty)}$,
one gets $\|\mathbb{WK}_d[f]\|_{\power^{\alpha-1}_\infty(0,\infty)} \leq \|f\|_{\power^{-\alpha}_\infty(0,\infty)}$.

\subsection{Stability of the integral gauges $\power_1$ and $\power_0$}
For the second part of the result, the method of variation of the constant in \eqref{ODE1} provides:
\begin{equation}
\frac{\mathbb{WK}_d[f](\lambda_1)}{\lambda_1^{\alpha-1}}
= \frac{\mathbb{WK}_d[f](\lambda_0)}{\lambda_0^{\alpha-1}} 
- \int_{\lambda_0}^{\lambda_1} \left(\int_0^\infty H_d(st) \varepsilon(s)f(s) ds\right)
\frac{dt}{t^\alpha}\cdotp
\end{equation}
Let us estimate the right-hand side by the change of variable $\sigma=st$:
\[
\int_{\lambda_0}^{\lambda_1} \left(\int_0^\infty H_d(st) |\varepsilon(s)| f(s) ds\right)\frac{dt}{t^\alpha} \leq
c_d^+\int_0^\infty \left(\int_{s\lambda_0}^{s\lambda_1} \frac{\sigma^{2-\alpha}}{\pi^{-2}+\sigma^2} d\sigma \right) s^{\alpha-1} |\varepsilon(s)| f(s) ds
\]
where $c_d^+\leq \frac{d}{d-1}$ is given by \eqref{kernel:intermidiary}.
For $\alpha\in(1,3)$, the inner integral is:
\[
\int_0^\infty \frac{\sigma^{2-\alpha}}{\pi^{-2}+\sigma^2} d\sigma =\frac{\pi^\alpha}{2\cos\left(\pi-\frac{\alpha \pi}{2}\right)}\cdotp
\]
Moreover, proposition~\ref{prop:powerlawconsequence} implies that $f(s)\leq c_0 s^{-\alpha} \exp(\|f\|_{\power_0^{-\alpha}(0,\infty)})$
with $c_0=\lim\limits_{x\to+\infty}  x^\alpha f(x)$.
According to~\eqref{ODEdefGauges}, one has $\int_0^\infty |\varepsilon(s)| \frac{ds}{s}=\|f\|_{\power_1^{-\alpha}(0,\infty)}$,
so the estimates boil down to:
\[
\left|\frac{\mathbb{WK}_d[f](\lambda_1)}{\lambda_1^{\alpha-1}} - \frac{\mathbb{WK}_d[f](\lambda_0)}{\lambda_0^{\alpha-1}}\right|
\leq \frac{c_d^+ \pi^\alpha c_0 }{2\cos\left(\pi-\frac{\alpha \pi}{2}\right)} 
\|f\|_{\power_1^{-\alpha}(0,\infty)} \exp(\|f\|_{\power_0^{-\alpha}(0,\infty)}).
\]
Similarly, using~\eqref{main}, \eqref{kernel:intermidiary} and proposition~\ref{prop:powerlawconsequence}, one gets a pointwise lower bound of $\mathbb{WK}_d[f]$:
\[
\frac{\mathbb{WK}_d[f](\lambda_2)}{\lambda_2^{\alpha-1}} 
\geq 
\frac{c_d^- \pi^\alpha c_0 }{2\cos\left(\pi-\frac{\alpha \pi}{2}\right)} 
\exp(-\|f\|_{\power_0^{-\alpha}(0,\infty)}).
\]
One thus gets for any triplet $\lambda_0$, $\lambda_1$, $\lambda_2>0$:
\begin{equation}
\left|\frac{\mathbb{WK}_d[f](\lambda_0)}{\lambda_0^{\alpha-1}} - \frac{\mathbb{WK}_d[f](\lambda_1)}{\lambda_1^{\alpha-1}}\right|
\leq \frac{c_d^+}{c_d^-} 
\|f\|_{\power_1^{-\alpha}(0,\infty)} \exp(2\|f\|_{\power_0^{-\alpha}(0,\infty)}) \cdot
\frac{\mathbb{WK}_d[f](\lambda_2)}{\lambda_2^{\alpha-1}}
\cdotp
\end{equation}
Next, one computes the gauge $\|f\|_{\power_0^{\alpha-1}(0,\infty)}$:
\[
\|f\|_{\power_1^{\alpha-1}(0,\infty)} = \sup_{\lambda,\lambda'} \left|
\log\frac{\mathbb{WK}_d[f](\lambda)}{\lambda^{\alpha-1}} - \log \frac{\mathbb{WK}_d[f](\lambda')}{(\lambda')^{\alpha-1}}
\right|.
\]
As $\log$ is a $c^{-1}$-Lipschitz function on $[c,\infty)$ with $c=\inf \frac{\mathbb{WK}_d[f](\lambda)}{\lambda^{\alpha-1}}>0$,
one immediately gets~\eqref{global2}.
\cqfd

\section{Universal regimes for the Wiener-Khinchin transform}
\label{proof:thm23}

Let us now focus on the two regimes $\lambda\to0$ and $\lambda\to+\infty$
for $\mathbb{WK}_d[f]$. The game is to find reasonable assumptions on $f$
that will lead to a universal behavior.

\subsection{Universal quadratic regime at the origin (proof of theorem~\ref{thm2})}\label{par:proofthm2}

In this section, we prove that $\mathbb{WK}_d[f](\lambda)\propto \lambda^2$ as $\lambda\to 0$.

\paragraph{Assumptions.}
In this section, let us assume that $f$ is smooth and satisfies:
\begin{gather}
\label{smallscale:assumption2}
\int_0^\infty (1+k^2) f(k) dk <\infty.
\end{gather}
In particular, one has
\[
\lim_{K\to\infty} \frac{K^2 \int_K^\infty f(k) dk}{\int_0^K k^2 f(k) dk} =0.
\]
Let us choose $\delta_0>0$ such that $\delta_0^2=\frac{d}{2\pi^2}<\frac{d+2}{2\pi^2}$.
For any $\delta\in (0,\delta_0]$, one defines $K_\delta^\sharp$ (which is expected to be large) as the smallest real number such that
\begin{subequations}
\begin{equation}\label{UR0A1}
\forall K\geq K_\delta^\sharp, \qquad 
K^2 \int_K^\infty f(k) dk \leq \frac{\delta^4 }{2} 
\int_0^K k^2 f(k) dk.
\end{equation}
In dimension $d\in\{1,2\}$, one also imposes that $K_\delta^\sharp$ satisfies: 
\begin{equation}\label{UR0A2}
\forall K\geq K_\delta^\sharp, \qquad 
\int_K^\infty k^2 f(k) dk \leq 
\frac{\delta^2}{2} \int_0^K k^2 f(k) dk.
\end{equation}
\end{subequations}
In this section, one will assume that $\delta\leq \delta_0$ and that
\[
0<\lambda\leq \frac{\delta}{K_\delta^\sharp}\cdotp
\]
\paragraph{Remark.} 
In general, one expects $K_\delta^\sharp\to+\infty$ as $\delta\to0$ unless $f$ is compactly supported.
When $d\geq 3$, the following proof will still hold if \eqref{smallscale:assumption2} is replaced by the weaker assumption:
\begin{equation}\label{UR0AltAss}
f\in L^1(\R_+;\R_+^\ast) \quad\text{and}\quad
\lim_{K\to\infty} \frac{K^2 \int_K^\infty f(k) dk}{\int_0^K k^2 f(k) dk} =0
\end{equation}
which holds not only under~\eqref{smallscale:assumption2} but also if $f(k) \sim c/k^3$ at infinity.
\paragraph{Proof  of theorem~\ref{thm2} in $\power^2_\infty$ gauge.}
One has the following decomposition~:
\begin{equation}
\mathbb{WK}_d[f](\lambda) - \frac{2\pi^2}{d}\cdot\lambda^2 \int_0^{\delta/\lambda} k^2 f(k)dk =
 \int_0^{\delta/\lambda} \left[ H_d(\lambda k) - \frac{2\pi^2}{d} (\lambda k)^2 \right] f(k)dk +  \int_{\delta/\lambda}^\infty H_d(\lambda k) f(k)dk.
 \end{equation}
By the kernel expansion at the origin~\eqref{kernel:origin}, the first integral on the right-hand side is negative and its absolute value is bounded by
\[
\frac{2\pi^4}{d(d+2)} \int_0^{\delta/\lambda}  (\lambda k)^4 f(k)dk \leq \frac{2\pi^4}{d(d+2)} \cdot \lambda^2 \delta^2 \int_0^{\delta/\lambda} k^2 f(k)dk.
\]
The second integral is positive. As $\frac{2\pi^2}{d}-\frac{2\pi^4}{d(d+2)}\cdot\delta_0^2  > \frac{\pi^2}{d} = \frac{1}{2\delta_0^2}$, one gets a lower bound:
\begin{equation}\label{UR0LB}
\mathbb{WK}_d[f](\lambda)\geq \frac{\lambda^2}{2\delta_0^2}\int_0^{\delta/\lambda} k^2 f(k)dk
\qquad\left(0<\lambda\leq \frac{\delta}{K_\delta^\sharp}\right).
\end{equation}
Estimate~\eqref{kernel:intermidiary} followed by~\eqref{UR0A1} with $K=\delta/\lambda\geq K_\delta$ provides a control of the the positive integral:
\[
\int_{\delta/\lambda}^\infty H_d(\lambda k) f(k)dk
\leq 
c_d^+  \int_{\delta/\lambda}^\infty f(k)dk \leq 
c_d^+ \cdot 
\frac{\delta^2 \lambda^2}{2} \int_0^{\delta/\lambda} k^2 f(k)dk.
\]
As $c_d^+\leq 2$, one gets an upper bound that is barely more than twice the lower bound~\eqref{UR0LB}:
\begin{equation}\label{UR0UB}
\mathbb{WK}_d[f](\lambda)\leq  \left(\delta_0^2+\frac{1}{\delta_0^2}\right) \lambda^2 \int_0^{\delta/\lambda} k^2 f(k)dk
\qquad\left(0<\lambda\leq \frac{\delta}{K_\delta^\sharp}\right).
\end{equation}

\medskip
The next step is to control the slope of $\mathbb{WK}_d[f]$ in log-log coordinates. The computation goes as follows:
\begin{equation}\label{UR0WKDecomp}
\lambda \mathbb{WK}_d[f]'(\lambda) - 2 \mathbb{WK}_d[f](\lambda) = -\int_0^\infty L_d(k\lambda) f(k) dk
\end{equation}
with $L_d(z)=2H_d(z)-z H_d'(z)$.
Using lemma~\ref{lemma:L} from the appendix and the lower bound \eqref{UR0LB}, one gets:
\[
\int_0^{\delta/\lambda} |L_d(k\lambda)| f(k) dk
\leq  C_L \int_0^{\delta/\lambda} (k\lambda )^4 f(k) dk
\leq C_L \delta^2 \lambda^2 \int_0^{\delta/\lambda} k^2 f(k) dk
\leq 2C_L \delta_0^2 \delta^2  \mathbb{WK}_d[f](\lambda).
\]
The control of the other half of the integral depends on the dimension.
\paragraph{Case $d\geq3$.} Using lemma~\ref{lemma:L} followed by~\eqref{UR0A1} with $K=\delta/\lambda\geq K_\delta$ and
\eqref{UR0LB},  one gets:
\[ 0\leq
\int_{\delta/\lambda}^\infty L_d(k\lambda) f(k) dk
\leq  C_L \int_{\delta/\lambda}^\infty f(k) dk
\leq \frac{C_L\delta^2 \lambda^2}{2} \int_0^{\delta/\lambda} k^2 f(k)dk
\leq C_L \delta_0^2 \delta^2  \mathbb{WK}_d[f](\lambda).
\]
Combining the two parts gives,
for $d\geq3$ and $\delta\leq \delta_0$:
\begin{equation}\label{UB0Cinf}
\|\mathbb{WK}_d[f]\|_{\power^2_\infty\big(0,\frac{\delta}{K_\delta^\sharp}\big)} \leq 
3 C_L   \delta_0^2 \cdot \delta^2.
\end{equation}
Note that in this case, the sign of $L_d$ implies that the log-log slope of $\mathbb{WK}_d[f](\lambda)$ will never
exceed the asymptotic slope $2$, \ie one has \eqref{upperslope}.

\paragraph{Case $d\leq2$.} When $d=2$, one uses  \eqref{UR0A1}-\eqref{UR0A2} and Cauchy-Schwarz to absorb the slow growth of the kernel:
\begin{align*}
\int_{\delta/\lambda}^\infty |L_d(k\lambda)| f(k) dk
&\leq  C_L \lambda^{1/2} \int_{\delta/\lambda}^\infty k^{1/2} f(k) dk
\leq C_L \lambda^{1/2}  \left(\int_{\delta/\lambda}^\infty k^{2} f(k) dk\right)^{1/4} \left(\int_{\delta/\lambda}^\infty f(k) dk\right)^{3/4}\\
&\leq \frac{C_L \delta^2 \lambda^2}{2} \int_0^{\delta/\lambda} k^2f(k) dk 
\leq  C_L \delta_0^2 \delta^2 \mathbb{WK}_d[f](\lambda).
\end{align*}
When $d=1$, the computation is similar:
\begin{align*}
\int_{\delta/\lambda}^\infty |L_d(k\lambda)| f(k) dk
&\leq  C_L \lambda \int_{\delta/\lambda}^\infty k f(k) dk
\leq C_L \lambda  \left(\int_{\delta/\lambda}^\infty k^{2} f(k) dk\right)^{1/2} \left(\int_{\delta/\lambda}^\infty f(k) dk\right)^{1/2}\\
&\leq \frac{C_L \delta^2 \lambda^2}{2} \int_0^{\delta/\lambda} k^2f(k) dk 
\leq  C_L \delta_0^2 \delta^2 \mathbb{WK}_d[f](\lambda).
\end{align*}
In both cases, one gets~\eqref{UB0Cinf} again.
Note however that this time, the sign-change of $L_d$ implies that the log-log slope of $\mathbb{WK}_d[f]$ might oscillate around the asymptotic slope $2$
for some functions $f$.

\paragraph{Proof  of theorem~\ref{thm2} in $\power^2_1$ gauge.}
By definition, for any $A>0$, one has
\begin{align*}
\| \mathbb{WK}_d[f]\|_{\power_1^2(0,A)}
& = \int_0^A \left| \frac{\lambda  \mathbb{WK}_d[f]'(\lambda)}{ \mathbb{WK}_d[f](\lambda)} -2\right| \frac{d\lambda}{\lambda} \\
& \leq \int_0^\infty  \left( \int_0^A \frac{|L_d(k \lambda)|}{\lambda \mathbb{WK}_d[f](\lambda)}  d\lambda \right) f(k) dk.
\end{align*}
Applying lemma~\ref{lemma:L}, one gets:
\[
\| \mathbb{WK}_d[f]\|_{\power_1^2(0,A)}
\leq C_L \int_0^\infty  \left( \int_0^A \frac{\lambda^3}{\mathbb{WK}_d[f](\lambda)}  d\lambda \right) k^4 f(k) dk.
\]
For $\delta\leq \delta_0$ and $A=\delta/K_{\delta}^\sharp$, the lower bound~\eqref{UR0LB} implies for $\lambda<A$:
\[
\mathbb{WK}_d[f](\lambda)\geq \frac{\lambda^2}{2\delta_0^2}\int_0^{K_{\delta}} k^2 f(k)dk.
\]
Combining the two estimates and $K_{\delta}^\sharp \geq K_{\delta_0}^\sharp \geq1$ provides (assuming the right-hand side is finite):
\begin{equation}\label{UR0C1}
\| \mathbb{WK}_d[f]\|_{\power_1^2\big(0,\frac{\delta}{K_{\delta}^\sharp}\big)}
\leq \frac{ \displaystyle C_L \delta_0^2 \int_0^\infty  k^4 f(k) dk}{\displaystyle \int_0^{K_{\delta_0}^\sharp} k^2 f(k)dk}
\left(\frac{\delta}{K_{\delta}^\sharp}\right)^2
\cdotp
\end{equation}
This is~\eqref{smallscale:conclusion2}. Note that if $K_\delta^\sharp \to \infty$ as $\delta\to0$, then the right-hand
side becomes smaller than $C_L\delta_0^2\delta^2$ for a small enough~$\delta$. The constant is then independent of $f$,
but the smallness threshold does depend on $f$.

\paragraph{Alternate estimate of the $\power^2_1$ gauge.}
To avoid using the $4^\text{th}$ momentum of $f$ in~\eqref{UR0C1}, one can also transform the $\power^2_\infty$ estimate into a $\power_1^2$ one.
Choosing $K_\delta\geq K_\delta^\sharp$ as a \textsl{smooth} decreasing function of $\delta$ (\ie $K_\delta$ increases when $\delta$ decreases),
the function $\Phi_f:\delta\mapsto\frac{\delta}{K_\delta}$ is sublinear and increasing on $[0,\delta_0]$.
In particular, this function admits an inverse function near the origin:
\[
\Phi_f^{-1}(\lambda) = \inf \left\{  \delta>0 \,;\, K_\delta \leq \frac{\delta}{\lambda}
\right\}.
\]
One has, using the previous results:
\begin{align*}
\| \mathbb{WK}_d[f]\|_{\power_1^2(0,A)}
&= \int_0^A \left| \frac{\lambda  \mathbb{WK}_d[f]'(\lambda)}{ \mathbb{WK}_d[f](\lambda)} -2\right| \frac{d\lambda}{\lambda}
\leq \int_0^A \|\mathbb{WK}_d[f]\|_{\power^2_\infty(0,\lambda)} \frac{d\lambda}{\lambda}\\
&\leq 3 C_L \delta_0^2 \int_0^A \frac{ \Phi_f^{-1}(\lambda)^2}{\lambda}d\lambda
= 3 C_L \delta_0^2 \int_0^{\Phi_f^{-1}(A)} \frac{\delta^2 \Phi_f'(\delta)}{\Phi_f(\delta)} d\delta.
\end{align*}
Applying this inequality with $A=\Phi_f(\delta)$ and cleaning up the derivative finally provides:
\begin{equation}\label{UR0C2}
\|\mathbb{WK}_d[f]\|_{\power_1^2(0,\frac{\delta}{K_\delta})} \leq
3 C_L \delta_0^2  \int_0^{\delta} s\left(1-\frac{s K'_s}{K_s}\right)ds.
\end{equation}
which offers an alternative to~\eqref{UR0C1} that does not require additional decay of $f$ beyond~\eqref{smallscale:assumption2}.
The right-hand side is $\mathcal{O}(\delta^2)$ if the log-log slope of $K_\delta$ is bounded as $\delta\to0$.

\medskip\noindent
This concludes the proof of Theorem~\ref{thm2}.
\cqfd

\subsection{Universal constant regime at infinity (proof of theorem~\ref{thm3})}\label{par:URinf}

In this section, we prove that $\mathbb{WK}_d[f](\lambda)\propto 1$ as $\lambda\to +\infty$.

\paragraph{Assumptions.}
In this section, let us assume that $d\geq2$ and that $f$ is smooth and satisfies:
\begin{equation}\label{largescale:assumption1}
\int_0^\infty f(k) dk <\infty.
\end{equation}
For any $\eta>0$, one defines $K_\eta^\flat$ (which is expected to be small) as the largest real number such that
\begin{subequations}
\begin{equation}\label{URinfA1}
\forall K\leq K_\eta^\flat, \qquad 
\int_0^K f(k) dk \leq \frac{1}{2} \:
\eta^{-\frac{d-1}{2}} \int_K^\infty f(k) dk.
\end{equation}
Additionally, if $d=2,3$, one may reduce the value of $K_\eta^\flat$ enough to ensure also that:
\begin{align}\label{URinfA2}
\sup_{\lambda\geq \eta / K_\eta^\flat} \left|\int_0
^\infty e^{2 i \pi \lambda k} f(k) dk \right|
&\leq \eta^{-1} \int_0^\infty f(k) dk
\qquad (\text{if }d=3),\\
\label{URinfA3}
\sup_{\lambda\geq \eta / K_\eta^\flat} \left|\int_0
^\infty \lambda k J_1(2 \pi \lambda k)  f(k) dk \right|
& \leq \eta^{-1/2} \int_0^\infty f(k) dk
\qquad (\text{if }d=2).
\end{align}
\end{subequations}
Observe that $\eta\mapsto K_\eta^\flat$ is a decreasing function because
\[
\frac{d}{dK}\left(\frac{\int_0^K f(k)dk}{\int_K^\infty f(k)dk}\right) = \left(\int_0^\infty f(k)dk\right)\left(\int_K^\infty f(k)dk\right)^{-2}f(K) \geq0.
\]
In general, $K_\eta^\flat$ tends to $0$ as $\eta\to\infty$, unless $f\equiv0$ near the origin.
Let us choose $\eta_0$ such that
\begin{equation}\label{defmu0}
\eta_0 \geq \max\left\{
\left[ 1+(1+\mathbf{1}_{d\geq4})\Gamma\left(\tfrac{d}{2}\right)\pi^{-d/2} \right]^{2/(d-1)};
\mathbf{1}_{d\geq4}\cdot c_d^0\cdot \frac{\pi^{d/2}}{\Gamma\left(\tfrac{d}{2}\right)}
\right\}.
\end{equation}
The constant $c_d^0$ is defined in the appendix as an upper bound of the remainder term of \eqref{kernel:asymptotic}.
In what follows, one will assume that $\eta\geq \eta_0$ and that
\[
\lambda\geq \frac{\eta}{K_\eta^\flat}\cdotp
\]

\paragraph{Proof of theorem~\ref{thm3}.}
For large $\lambda$, one expects  $\mathbb{WK}_d[f](\lambda)$ to be roughly constant.
The first step of the proof relies on the following decomposition:
\begin{equation}
\mathbb{WK}_d[f](\lambda) - \int_0^\infty f(k)dk = \int_{\eta/\lambda}^\infty (H_d(\lambda k)-1) f(k)dk
+ \int_0^{\eta/\lambda} H_d(\lambda k)f(k) dk
-  \int_0^{\eta/\lambda} f(k) dk.
\end{equation}
For the first integral, one uses \eqref{kernel:Hminus1}, which holds because $\lambda k\geq \eta \geq \eta_0 >\mathbf{1}_{d\geq4}\cdot c_d^0\cdot \pi^{d/2}/\Gamma\left(\tfrac{d}{2}\right)$:
\[
\left| \int_{\eta/\lambda}^\infty (H_d(\lambda k)-1) f(k)dk \right|
\leq (1+\mathbf{1}_{d\geq4})\Gamma\left(\tfrac{d}{2}\right)\pi^{-d/2} \cdot \eta^{-\frac{d-1}{2}} \int_{\eta/\lambda}^\infty f(k)dk.
\]
The second integral can be reduced to the third one by using~\eqref{kernel:intermidiary} and $c_d^+\leq 2$:
\[
\left|\int_0^{\eta/\lambda} H_d(\lambda k)f(k) dk\right| \leq  2\int_0^{\eta/\lambda} f(k) dk.
\]
Finally, the third one is dealt with by using \eqref{URinfA1} with $K=\eta/\lambda\leq K_\eta^\flat$:
\[
\int_0^{\eta/\lambda} f(k) dk \leq
\frac{1}{2}\: \eta^{-\frac{d-1}{2}} \int_{\eta/\lambda}^\infty f(k) dk.
\]
Combining the three estimates gives:
\[
\left|\mathbb{WK}_d[f](\lambda) - \int_0^\infty f(k)dk \right| \leq
\left[1+(1+\mathbf{1}_{d\geq4})\Gamma\left(\tfrac{d}{2}\right)\pi^{-d/2}\right] \cdot \eta^{-\frac{d-1}{2}} \int_{\eta/\lambda}^\infty f(k) dk.
\]
Note that, due to the sign change between the second and third term, only one of them contributes
to each upper or lower estimate, which slightly improves the constant.
Using the definition of $\eta_0$,  one has thus established the following estimate, which provides both an upper
and a lower bound of~$\mathbb{WK}_d[f]$:
\begin{equation}\label{limWKd}
\left|\mathbb{WK}_d[f](\lambda) - \int_0^\infty f(k)dk \right| \leq \left(\frac{\eta}{\eta_0}\right)^{-\frac{d-1}{2}}
\int_{\eta/\lambda}^\infty f(k)dk
\qquad\left(\lambda\geq  \frac{\eta}{K_\eta^\flat}\right).
\end{equation}

\bigskip
Next, one shows the smallness of the slope in log-log coordinates using the following expression:
\begin{equation}\label{URinfWKDecomp}
\lambda \mathbb{WK}_d[f]'(\lambda) = -\int_0^\infty (\lambda k) H_d'(\lambda k) f(k) dk.
\end{equation}
The goal is to compute the rate at which the slope stabilises,
which depends on the dimension.
Using lemma~\ref{lemma:L2}, one gets for $d\geq3$:
\[
\left| \int_0^{\eta/\lambda} (\lambda k) H_d'(\lambda k) f(k) dk\right| \leq 
2 \left|\int_0^{\eta/\lambda} H_d(\lambda k)f(k) dk\right|
\leq  \eta^{-\frac{d-1}{2}} 
\int_{\eta/\lambda}^\infty f(k) dk,
\]
while for $d=2$, one gets only:
\[
\left| \int_0^{\eta/\lambda} (\lambda k) H_d'(\lambda k) f(k) dk\right| \leq 
2(1+\sqrt{\eta}) \left|\int_0^{\eta/\lambda} H_d(\lambda k)f(k) dk\right|
\leq C\int_{\eta/\lambda}^\infty f(k) dk.
\]
The second half of the integral requires further attention.

\paragraph{Case $d\geq4$.}
The second estimate in lemma~\ref{lemma:L2} gives:
\[
\left| \int_{\eta/\lambda}^\infty (\lambda k) H_d'(\lambda k) f(k) dk\right| \leq C_d \eta^{-\frac{d-3}{2}} \int_{\eta/\lambda}^\infty H_d(\lambda k)f(k) dk,
\]
which is a slower decay rate than $\eta^{-(d-1)/2}$, by an order of magnitude. One thus gets:
\begin{equation}\label{URinfCd4}
\|\mathbb{WK}_d[f]\|_{\power^0_\infty(\frac{\eta}{K_\eta^\flat},\infty)} =\mathcal{O}(\eta^{-\frac{d-3}{2}})
\end{equation}
with a universal numerical constant that depends only on the dimension.

\paragraph{Remark.}
For $d\geq 3$, lemma~\ref{lemma:L} ensures that $z |H_d'(z)|\leq 2H_d(z)$. The decomposition~\eqref{URinfWKDecomp}
thus provides a pointwise bound
$\left| \lambda \mathbb{WK}_d[f]'(\lambda) \right| \leq 2 \mathbb{WK}_d[f](\lambda)$, \ie
\begin{equation}\label{URinfC0}
\|\mathbb{WK}_d[f]\|_{\power^0_\infty(0,\infty)} \leq2.
\end{equation}
This is a slight improvement over~\eqref{upperslope}. 
However,~\eqref{URinfCd4} and \eqref{URinfC0} are not sufficient to achieve an asymptotic stabilization of the slope of $\mathbb{WK}_d[f]$ when $d=2$ or $d=3$.

\paragraph{Case $d=3$.}
Let us compute the kernel of~\eqref{URinfWKDecomp} explicitly:
\[
z H'_3(z) = -\cos(2\pi z) + \frac{\sin(2\pi z)}{2\pi z}.
\]
As before, one has $\left| \int_0^{\eta/\lambda} (\lambda k) H_d'(\lambda k) f(k) dk\right| = \mathcal{O}(\eta^{-1}\cdot \mathbb{WK}_d[f](\lambda))$.
The next good term is:
\[
\left| \int_{\eta/\lambda}^\infty \frac{\sin (2\pi \lambda k)}{2\pi\lambda k} f(k) dk\right| \leq \frac{\eta^{-1}}{2\pi} \int_{\eta/\lambda}^\infty f(k) dk.
\]
For the last term, one uses the additional assumption~\eqref{URinfA2}, which ensures that
\[
\left|\int_0^\infty \cos(2\pi\lambda k) f(k) dk\right| \leq \frac{C}{\eta} \int_0^\infty f(k) dk
\]
for $\lambda\geq\eta/{K^\flat_\eta}$.
One has therefore established that:
\begin{equation}\label{URinfCd3}
\|\mathbb{WK}_3[f]\|_{\power^0_\infty(\frac{\eta}{K_\eta^\flat},\infty)} =\mathcal{O}(\eta^{-1}).
\end{equation}

\paragraph{Case $d=2$.}
One uses instead:
\[
zH'_2(z)=2\pi z J_1(2\pi z) = -2\sqrt{z} \cos\left(2\pi z+\frac{\pi}{4}+\mathcal{O}(z^{-2})\right) + \mathcal{O}(z^{-1/2}).
\]
Assumption~\eqref{URinfA3} ensures that:
\begin{equation}\label{URinfCd2}
\|\mathbb{WK}_2[f]\|_{\power^0_\infty(\frac{\eta}{K_\eta^\flat},\infty)} =\mathcal{O}(\eta^{-1/2}).
\end{equation}
This concludes the proof of theorem~\ref{thm3}.
\cqfd\par

%

\paragraph{Alternate control of the  $\power^0_\infty$ gauge when $d\geq2$.}
The hypergeometric function
\[
G_d(z) = \frac{4\pi^2}{3d} z^3 \cdot {}_1F_2\left(\frac{3}{2}; \frac{5}{2}, 1+\frac{d}{2} ; -\pi^2 z^2 \right) = \mathcal{O}(z^{\frac{3-d}{2}})
\]
is a primitive of $G_d'(z) = z H_d'(z)$.
An integration by part provides an alternate upper bound for $\lambda\geq \eta/K_\eta^\flat$:
\[
\lambda |\mathbb{WK}_d[f]'(\lambda)| 
= \left| \int_0^\infty  \frac{G_d(\lambda k)}{\lambda} f'(k) dk \right|
\leq C  \left(
\frac{\int_0^\infty |f'(k)| k^{-\frac{d-3}{2}}dk}{\int_0^\infty f(k) dk}
\right) \eta^{-\frac{d-1}{2}} |\mathbb{WK}_2[f](\lambda)|
\]
and thus
\begin{equation}\label{URinfCd4bis}
\|\mathbb{WK}_d[f]\|_{\power^0_\infty(\frac{\eta}{K_\eta^\flat},\infty)} 
\leq C  \left(
\frac{\int_0^\infty |f'(k)| k^{-\frac{d-3}{2}}dk}{\int_0^\infty f(k) dk}
\right) \eta^{-\frac{d-1}{2}}.
\end{equation}
In particular, one has:
\begin{align}\label{URinfCd2bis}
\|\mathbb{WK}_2[f]\|_{\power^0_\infty(\frac{\eta}{K_\eta^\flat},\infty)} 
&\leq C  \left(
\frac{\int_0^\infty k^{1/2} |f'(k)| dk}{\int_0^\infty f(k) dk}
\right) \eta^{-1/2},
\\
\label{URinfCd3bis}
\|\mathbb{WK}_3[f]\|_{\power^0_\infty(\frac{\eta}{K_\eta^\flat},\infty)} 
&\leq C  \left(
\frac{\int_0^\infty |f'(k)| dk}{\int_0^\infty f(k) dk}
\right) \eta^{-1}.
\end{align}
In dimensions 2 and 3, the decay rates of~\eqref{URinfCd2bis} and \eqref{URinfCd3bis} are respectively
the same as those of~\eqref{URinfCd2},~\eqref{URinfCd3}.
For $d\geq4$, the decay rate of \eqref{URinfCd4bis} exceeds that of \eqref{URinfCd4} by one order of magnitude.
Note, however, that the constant is not independent of $f$ anymore (or equivalently, if we want a universal constant,
then the threshold value of $\eta_0$ will depend on $f$)
and that some smoothness (a controlled growth of the derivative) of $f$ is required.

\paragraph{Remark on the case $d=1$.}
This case is not included in the statement of theorem~\ref{thm3} but one can use a slight variant of the integration by part technique to get a $\mathcal{O}(\eta^{-1})$ bound. One has:
\[
z H_1'(z) = 2\pi z \sin(2\pi z).
\]
Its primitive $G_1(z)=-z\cos(2\pi z) +\frac{1}{2\pi}\sin(2\pi z)$ is $\mathcal{O}(z)$ but
\[
\widetilde{G}_1(z) = -\frac{\cos(2\pi z) +\pi z \sin(2\pi z)}{2\pi^2} = \mathcal{O}(z)
\]
satisfies $\widetilde{G}_1''(z) = z H_1'(z)$ and thus, after two integrations by part,
\[
\lambda |\mathbb{WK}_1[f]'(\lambda)| = \left| \int_0^\infty 
\frac{\widetilde{G}_1(\lambda k)}{\lambda^2} f''(k) dk \right|
\leq C  \left(
\frac{\int_0^\infty |f''(k)| k dk}{\int_0^\infty f(k) dk}
\right) \eta^{-1} \cdot |\mathbb{WK}_1[f](\lambda)|
\]
\ie
\begin{equation}\label{URinfCd1bis}
\|\mathbb{WK}_1[f]\|_{\power^0_\infty(\frac{\eta}{K_\eta^\flat},\infty)} 
\leq C  \left(
\frac{\int_0^\infty |f''(k)| k dk}{\int_0^\infty f(k) dk}
\right) \eta^{-1}.
\end{equation}
Note that the fact that $G_1(z)$ and $\widetilde{G}_1(z)$ have the same order of magnitude is
specific to dimension~1. In general, the second order primitive $\widetilde{G}''_d(z) = z H'_d(z)$ satisfies $\widetilde{G}_d(z)=\mathcal{O}(z^{\frac{5-d}{2}})$
and one does not expect an improvement of~\eqref{URinfCd4bis} when $d\geq 2$.

\paragraph{Control of the $\power_1^0$ gauge.}
Similarly to the last remark of \S\ref{par:proofthm2},
one can also transform the $\power^0_\infty$ estimate into a $\power^0_1$ one.
Choosing $K_\eta\leq K_\eta^\flat$ as a smooth decreasing function of $\eta$,
the function $\Psi_f:\eta\mapsto\frac{\eta}{K_\eta}$ is increasing on $[\eta_0,\infty)$.
In particular, this function admits an inverse function at infinity:
\[
\Psi_f^{-1}(\lambda) = \sup \left\{  \eta>0 \,;\, K_\eta \geq \frac{\eta}{\lambda}
\right\}.
\]
One has, using \eqref{URinfCd4bis}:
\begin{align*}
\| \mathbb{WK}_d[f]\|_{\power_1^0(B,\infty)}
&= \int_B^\infty \left| \frac{\lambda  \mathbb{WK}_d[f]'(\lambda)}{ \mathbb{WK}_d[f](\lambda)}\right| \frac{d\lambda}{\lambda}
\leq \int_B^\infty \|\mathbb{WK}_d[f]\|_{\power^0_\infty(0,\lambda)} \frac{d\lambda}{\lambda}\\
&\leq C \left(
\frac{\int_0^\infty |f'(k)| k^{-\frac{d-3}{2}}dk}{\int_0^\infty f(k) dk}
\right)
\int_B^\infty \frac{d\lambda}{\lambda \Psi_f^{-1}(\lambda)^{\frac{d-1}{2}}}
\cdotp
\end{align*}
In particular, after a change of variables, one gets for $d\geq2$:
\begin{equation}\label{CHP1}
\| \mathbb{WK}_d[f]\|_{\power_1^0(\frac{\eta}{K_\eta},\infty)} \leq C \left(
\frac{\int_0^\infty |f'(k)| k^{-\frac{d-3}{2}}dk}{\int_0^\infty f(k) dk}
\right)
\int_{\eta}^\infty \left(1-\frac{s K_s'}{K_s}\right) \frac{ds}{s^{\frac{d+1}{2}}}\cdotp
\end{equation}

\paragraph{Remark.}
For example, when $f(k)$ is homogeneously asymptotic to $k^\beta$ near the origin
(which is the case in the hydrodynamics applications with \eg $\beta=2$), then one has:
\[
K_\eta^\flat \simeq \eta^{-\frac{d-1}{2(\beta+1)}} \ell^{\frac{1}{\beta+1}} \qquad\text{where}\qquad
\ell = \lim\limits_{\lambda\to\infty} \mathbb{WK}_d[f](\lambda) = \int_0^\infty f(k) dk.
\]
The inequality \eqref{CHP1} then reads $\| \mathbb{WK}_d[f]\|_{\power_1^0(\frac{\eta}{K_\eta^\flat},\infty)} = \mathcal{O}(\eta^{-\frac{d-1}{2}})$ \ie
\[
\| \mathbb{WK}_d[f]\|_{\power_1^0(B,\infty)} = \mathcal{O}\left(B^{-1/(\frac{2}{d-1}+\frac{1}{\beta+1})}\right)
\]
 and in particular,
as $\power_1^0$ controls $\power_0^0$ through~\eqref{gauges_inequality}:
\begin{equation}\label{CHP1bis}
\left| \mathbb{WK}_d[f](\lambda) - \ell \right| \sim 
\left| \ell \log \frac{\mathbb{WK}_d[f](\lambda)}{\ell} \right|
\lesssim \| \mathbb{WK}_d[f]\|_{\power_0^0(\lambda,\infty)}
= \mathcal{O}\left( \lambda^{-1/(\frac{2}{d-1}+\frac{1}{\beta+1})} \right).
\end{equation}
This convergence rate is exactly the same as the one given by \eqref{limWKd}.
However, as this estimate is deduced from a $\power_\infty^0$ control, \ie uniform bounds,
it does not take into account the possible cancellations that could be caused by the oscillations 
of $\mathbb{WK}_d[f]$ around its limit value.
When applied to $f=E(k)$ where $E$ is the energy spectrum~\eqref{energyspectrum} of an $L^2(\R^3)$ flow,
it provides an upper bound of the scalar correlation function~\eqref{S2Gamma}, namely
 $\mathcal{R}(\lambda)\lesssim \lambda^{-3/4}$ at infinity  for a generic Saffman spectrum, \ie $\beta=2$, and 
 $\mathcal{R}(\lambda)\lesssim \lambda^{-5/6}$ for a Batchelor spectrum, \ie $\beta=4$.

 \medskip
If one assumes additionally that $u\in L^1(\R^3)$, then this bound is not as good as the integral decay rate~\eqref{univConstFluids},
which translates loosely as $\mathcal{R}(\lambda)=\mathfrak{o}(\lambda^{-3})$.
But this is because the $L^1$ assumption on $u$ translates as \textsl{regularity} on the energy spectrum, \ie fewer fluctuations of $f$ than 
what could happen for the worst $L^2$ flow, which, according to~\eqref{largescale:assumption1}, is the level of generality of this section.

\begin{figure}[H]
\captionsetup{width=.85\linewidth}
\begin{center}
\includegraphics[width=.85\linewidth]{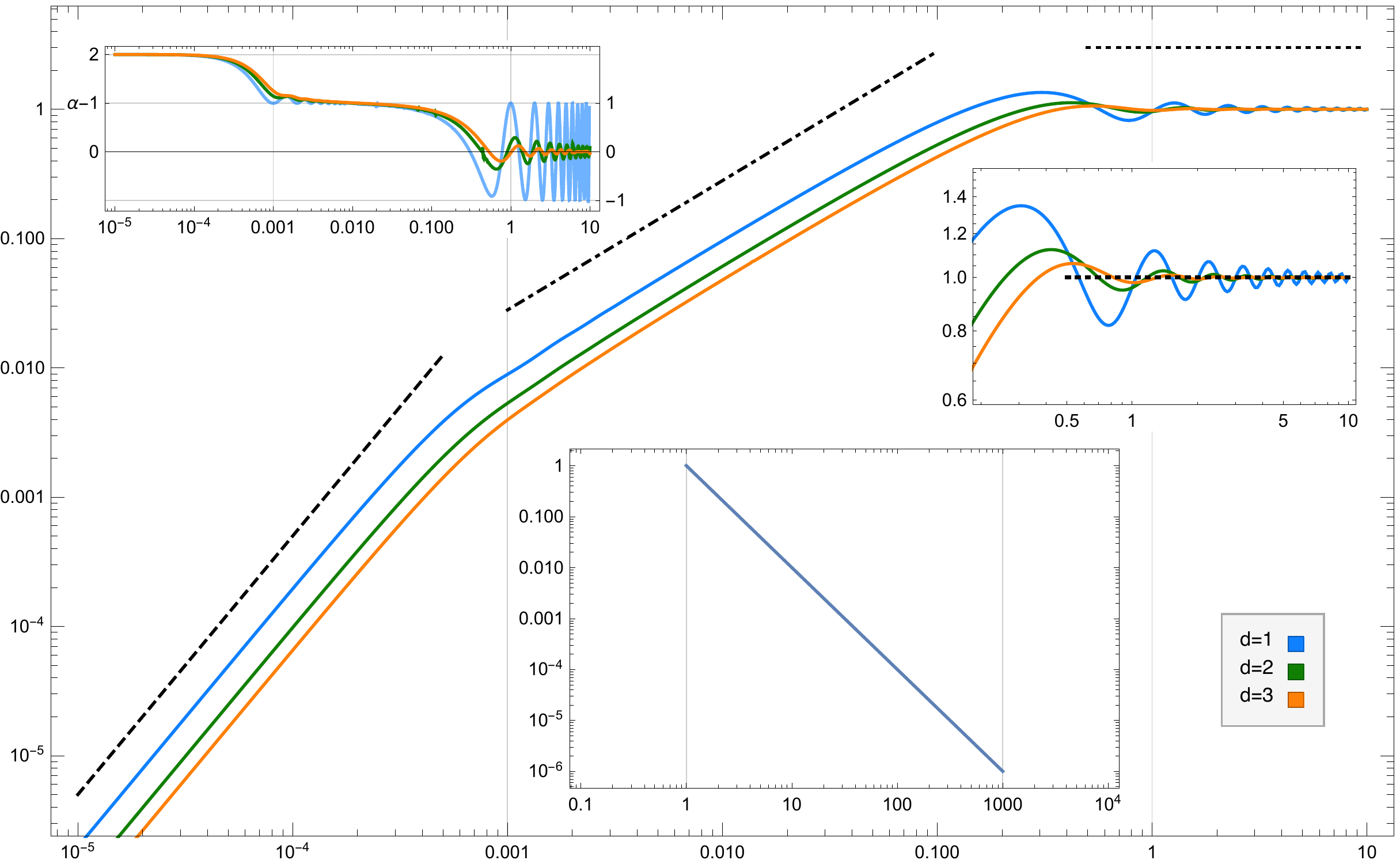}
\caption{\label{fig:compactsupport}\small 
Wiener-Khinchin transforms of $f(k)=k^{-\alpha} \mathbf{1}_{[k_1,k_2]}$.
The reference slopes are $2$, $\alpha-1$ and $0$.
The vertical delimiters mark the intervals~$[k_1,k_2]$ for $f$,
and $[k_1^{-1},k_2^{-1}]$ for $\mathbb{WK}_d[f]$.
The graph of~$f$ is in the bottom inset.
The top-right inset is a zoom on $\mathbb{WK}_d[f](\lambda)$ for large $\lambda$,
while the top-left inset displays the log-log slope of $\mathbb{WK}_d[f]$.
Note the persistence of high slopes in dimension~$d=1$, which is due to the discontinuity in $f$ (so \eqref{URinfCd1bis} does not apply).
}
\end{center}
\end{figure}

\begin{figure}[H]
\captionsetup{width=.85\linewidth}
\begin{center}
\includegraphics[width=.85\linewidth]{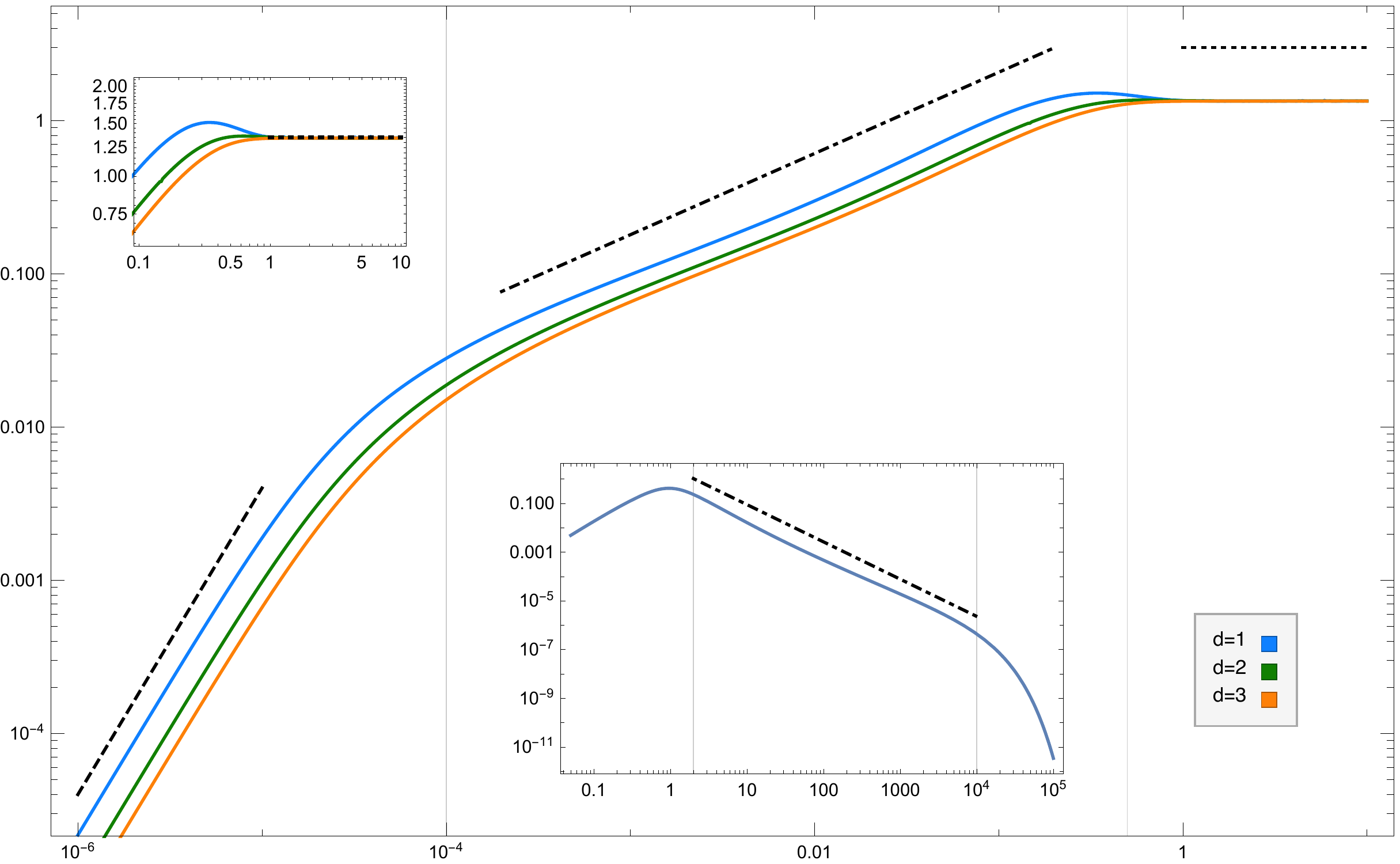}
\caption{\label{fig:threeregimefnct}\small
Wiener-Khinchin transforms of a three-regime functions \eqref{threeregimefnct}.
The layout is the same as in figure~\ref{fig:tworegimefnct}.
The reference slopes are 0, $0.53$ and 2 on the graph of~$\mathbb{WK}_d[f]$ and $-1.53$ on the graph of~$f$.
The marked intervals are dual from each other.}
\end{center}
\end{figure}

\begin{figure}[H]
\captionsetup{width=.85\linewidth}
\begin{center}
\includegraphics[width=.85\linewidth]{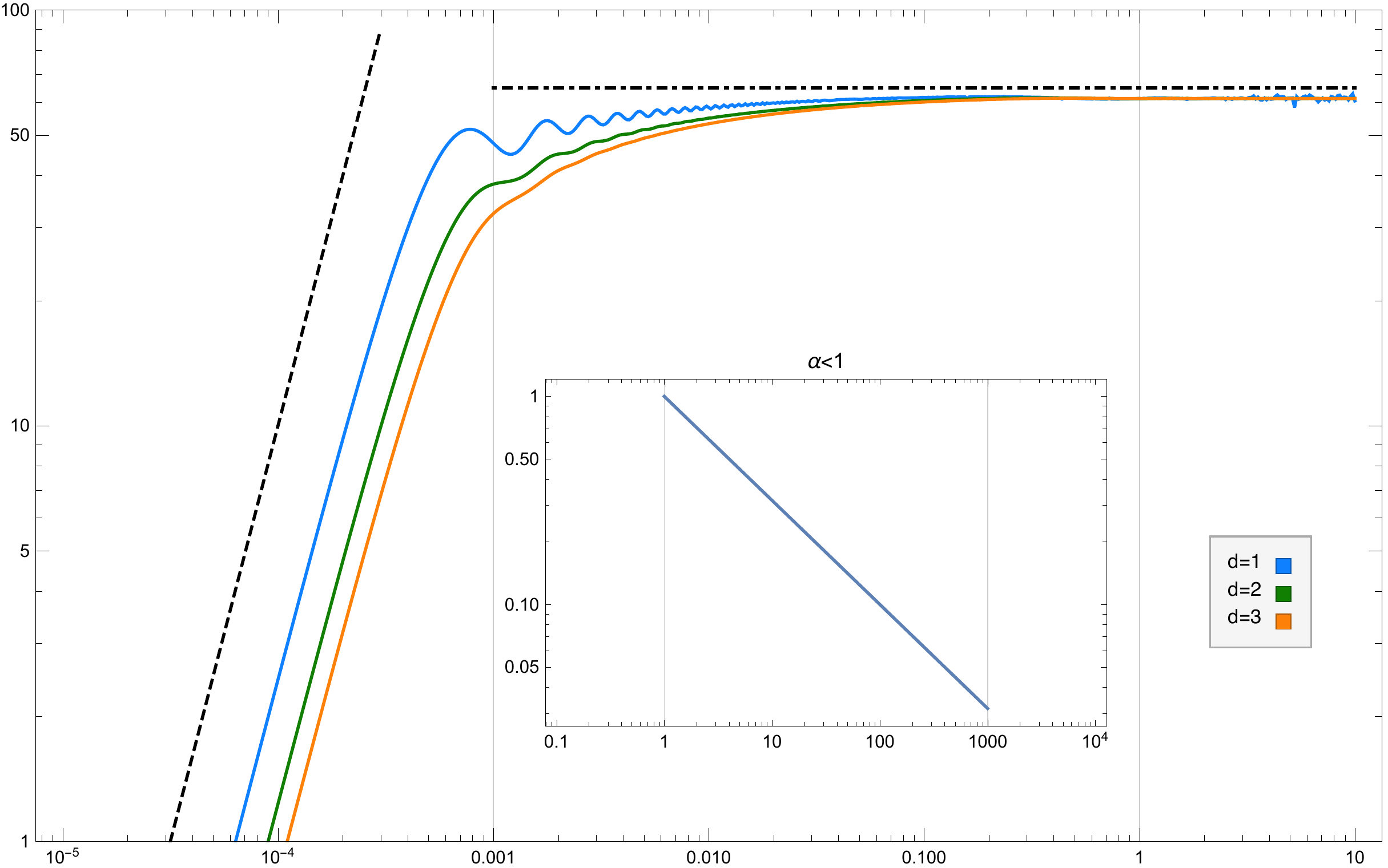}
\caption{\label{fig:smallalpha}\small
Wiener-Khinchin transforms of $f(k)=k^{-\alpha} \mathbf{1}_{[k_1,k_2]}$ with $\alpha<1$.
The reference slopes are 2 and 0. The inset is the graph of $f$.
One should disregard the numerical noise that occurs in dimension $d=1$ for $\lambda\gg1$.}
\end{center}
\end{figure}

\begin{figure}[H]
\captionsetup{width=.85\linewidth}
\begin{center}
\includegraphics[width=.85\linewidth]{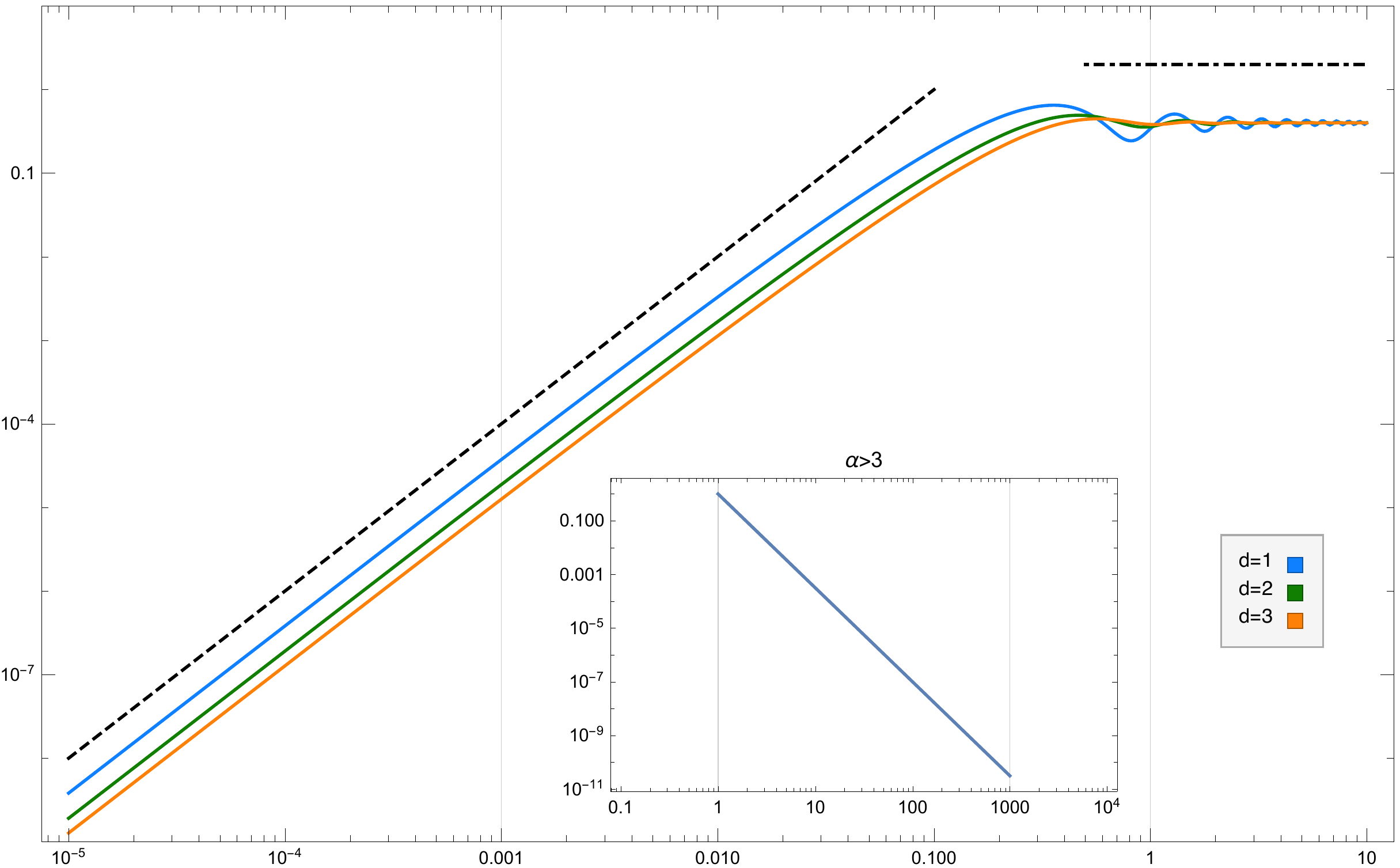}
\caption{\label{fig:largealpha}\small
Wiener-Khinchin transforms of $f(k)=k^{-\alpha} \mathbf{1}_{[k_1,k_2]}$ with $\alpha>3$.
The reference slopes are 2 and 0. The inset is the graph of $f$.}
\end{center}
\end{figure}

\begin{figure}[H]
\captionsetup{width=.85\linewidth}
\begin{center}
\includegraphics[width=.85\linewidth]{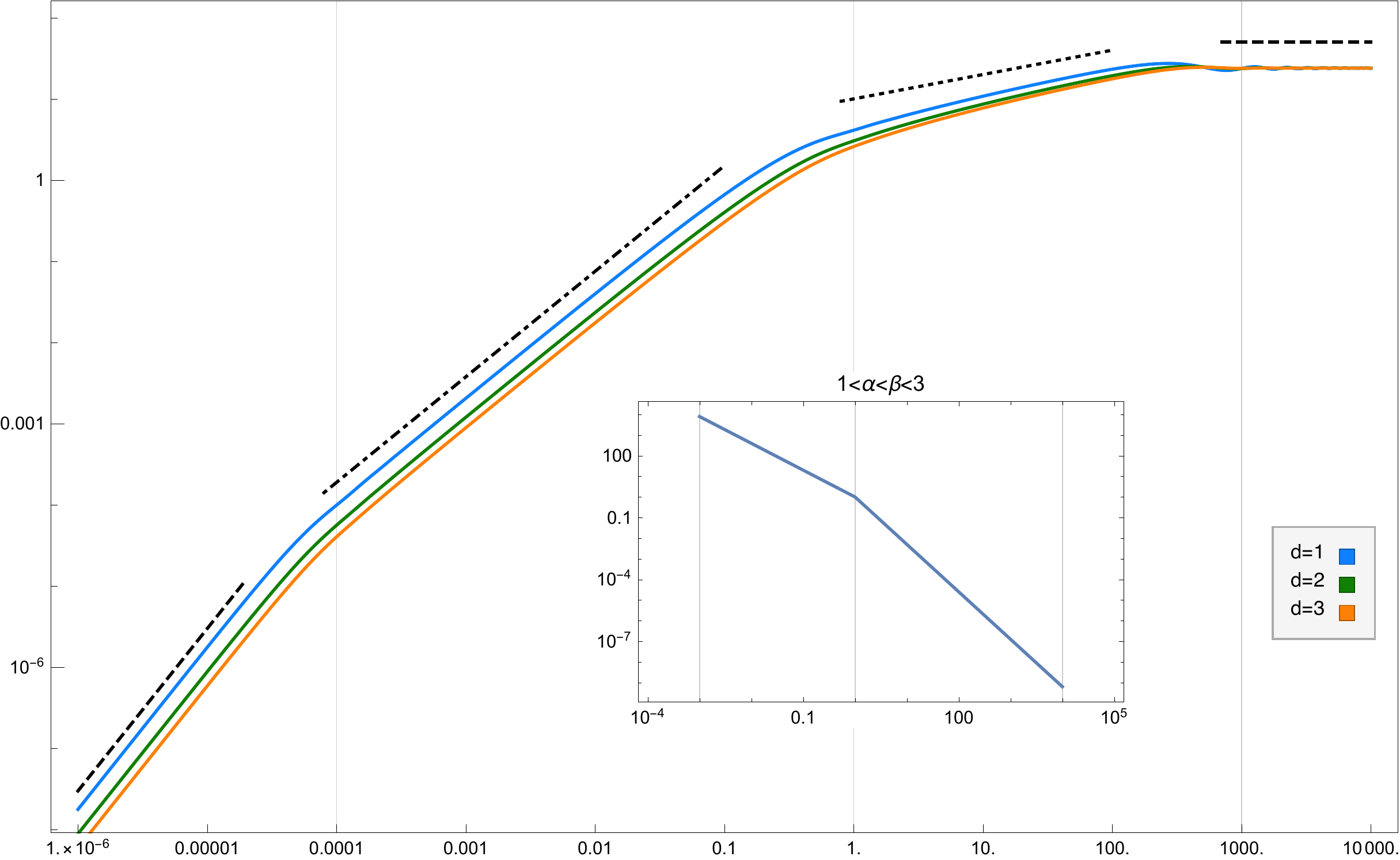}
\caption{\label{fig:fourregimefnct1}\small
Wiener-Khinchin transforms of a multi-regime function $f^{\alpha,\beta}_{k_1,k_2,k_3}$
defined by~\eqref{multiregimefnct} in the concave case $1<\alpha<\beta<3$.
From left to right, the reference slopes are 2, $\beta-1$, $\alpha-1$ and~0.}
\end{center}
\end{figure}

\begin{figure}[H]
\captionsetup{width=.85\linewidth}
\begin{center}
\includegraphics[width=.85\linewidth]{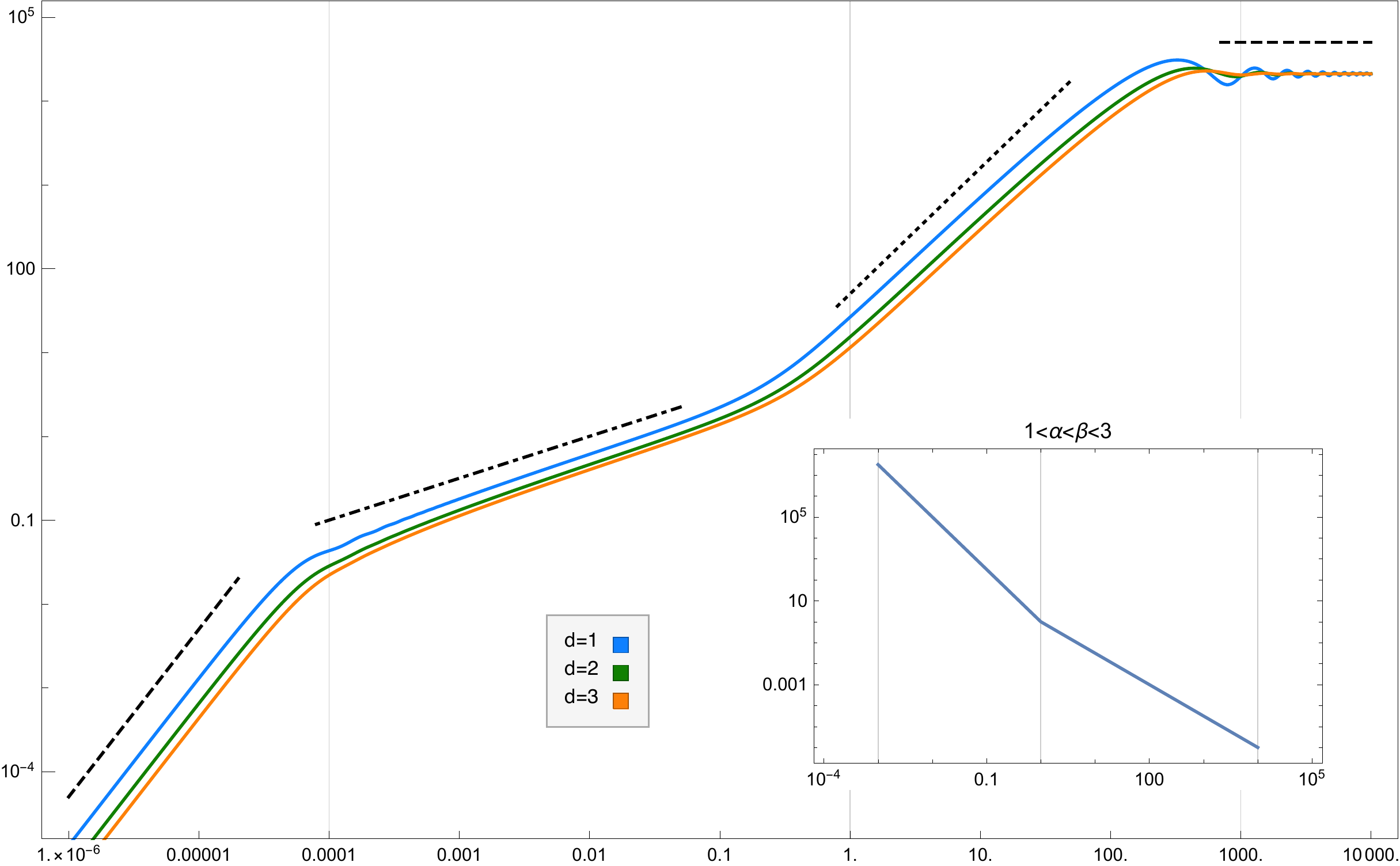}
\caption{\label{fig:fourregimefnct2}\small
Wiener-Khinchin transforms of a multi-regime function  $f^{\alpha,\beta}_{k_1,k_2,k_3}$
defined by~\eqref{multiregimefnct} in the convex case $1<\beta<\alpha<3$.
The reference slopes are 2, $\beta-1$, $\alpha-1$ and 0.}
\end{center}
\end{figure}

Note that a comparison between the figures~\ref{fig:compactsupport} and~\ref{fig:threeregimefnct} also suggests why the previous bound 
improves so slowly when $\beta$ increases. A better localization of the flow is concomitant with an increase of $\beta$  (see the example
on~p.\pageref{ex:localisationofflow}), which should improve the decay of $\mathcal{R}$. However, for a function $f$ supported
away from the origin (not that $f$ might then not be the energy spectrum of any realistic flow),
one has $\beta=\infty$ and the top-right and top-left insets of figure~\ref{fig:compactsupport} show
high oscillations of its $\mathbb{WK}_d$ transform, typical of a ringing artefact associated with a discontinuity.

In other words, the integrability of a flow (\ie the finiteness of its momentum) is not easily read on its energy spectrum,
which comes as no surprise as the Fourier transform does not play well with $L^p$ when $p\neq 2$.

\subsection{Comparison betwen $K_\eta^\flat$ and $K_\delta^\sharp$}

Up to now, one has shown that $ \mathbb{WK}_d[f](\lambda)$ behaves as $\lambda^2$ when $\lambda\ll \delta/K^\sharp_\delta$
and behaves like a constant when $\lambda\gg \eta/K^\flat_\eta$
(with improving constants when $\delta\to0$ or $\eta\to\infty$).
Let us check that
\begin{equation}\label{KflatKSharp}
K^\flat_\eta < K^\sharp_\delta.
\end{equation}
This inequality ensures that there is ``room'' in-between the two universal asymptotic regimes.
This intermediary range will subsequently be in the focus of theorem~\ref{thm4}.

\paragraph{Proof of~\eqref{KflatKSharp}.}
If~\eqref{KflatKSharp} does not hold, then, by definition, for any $K\in [K^\sharp_\delta,K^\flat_\eta]$, one has:
\[
\int_K^\infty f(k) dk \leq \frac{\delta^4}{2K^2} \int_0^K k^2 f(k) dk \leq \frac{\delta^4}{2} \int_0^K f(k) dk
\]
and
\[
\int_0^K f(k) dk \leq  \frac{1}{2}\, \eta^{-\frac{d-1}{2}} \int_K^\infty f(k) dk
\]
which combines into:
\[
1\leq \frac{1}{4} \, \delta^4\eta^{-\frac{d-1}{2}}.
\]
As $\delta_0^4 \eta_0^{-\frac{d-1}{2}} \leq \frac{d^2}{4\pi^4}
\left[ 1+(1+\mathbf{1}_{d\geq4})\Gamma\left(\tfrac{d}{2}\right)\pi^{-d/2} \right]^{-1} \leq 1$, one gets:
\[
4\leq \left(\frac{\eta}{\eta_0}\right)^{-\frac{d-1}{2}} \left(\frac{\delta}{\delta_0}\right)^4
\cdotp
\]
In particular, for $\delta\leq \delta_0$ and $\eta\geq\eta_0$, this is impossible
and thus $K^\flat_\eta < K^\sharp_\delta$.
\cqfd

\section{Duality of quasi-power-laws on finite ranges (proof of theorem~\ref{thm4})}
\label{proof:thm4}

The game is now to assume that $f$ is a quasi-power-law on a finite range $[k_1,k_2]$; we need to find the proper assumptions
on the tails that will lead to a quasi-power-law behavior
of $\mathbb{WK}_d[f]$ on the intermediary range $\frac{1}{k_2}\ll \lambda\ll \frac{1}{k_1}$.
To avoid confusion with the previous results, we introduce new names for the parameters ($\varepsilon$, small and $\mu$, large),
and we focus our study on the interval $\lambda\in [\frac{\mu}{k_2},\frac{\varepsilon}{k_1}]$.

\paragraph{Assumptions.}
Let us assume that $d\geq1$ and that $f$ is smooth and positive, and  that it satisfies:
\begin{equation}\label{assumption:tail}
\int_0^{k_1} k^2f(k) dk \leq C_1 k_1^3 f(k_1)
\qquad\text{and}\qquad
\int_{k_2}^\infty f(k)dk  \leq C_2 k_2 f(k_2).
\end{equation}
In dimensions $d\in\{1,2\}$, one requires additionally that:
\begin{equation}\label{assumption:dim2}
\int_{k_2}^\infty k |f'(k)|dk 
\leq C_2 k_2 f(k_2).
\end{equation}
One also assumes that the parameters $\varepsilon$ and $\mu$ satisfy:
\begin{equation}\label{epssmallmularge}
\sigma_\alpha(\varepsilon,\mu) := (\alpha-1) (\pi\varepsilon)^{3-\alpha} + (3-\alpha) (\pi \mu)^{-(\alpha-1)} \leq 1.
\end{equation}
One assumes finally that $\frac{k_2}{k_1} > \frac{\mu}{\varepsilon}$ to ensure that the interval $[\frac{\mu}{k_2},\frac{\varepsilon}{k_1}]$ is not
empty.

\paragraph{Proof of theorem~\ref{thm4}.}

After an integration by part, the following decomposition holds for any $k_1<k_2$:
\begin{equation}\label{HRdecomp}
\begin{aligned}
\lambda \mathbb{WK}_d[f]'(\lambda)-(\alpha-1)\mathbb{WK}_d[f](\lambda) =
&-\int_{k_1}^{k_2} H_d(\lambda k) \left[k f'(k)+\alpha f(k) \right] dk\\
&+H_d(\lambda k_2) k_2 f(k_2) - H_d(\lambda k_1) k_1 f(k_1)\\
&+\int_0^{k_1} \left[\lambda k H_d'(\lambda k)-(\alpha-1)H_d(\lambda k)\right] f(k)dk\\
&+\int_{k_2}^\infty \left[\lambda k H_d'(\lambda k)-(\alpha-1)H_d(\lambda k)\right] f(k)dk.
\end{aligned}
\end{equation}
The positivity of the kernel $H_d$ ensures that:
\begin{equation}
\left|\int_{k_1}^{k_2} H_d(\lambda k) \left[k f'(k)+\alpha f(k) \right] dk\right|
\leq \|f\|_{\power^{-\alpha}_\infty(k_1,k_2)} \mathbb{WK}_d[f](\lambda).
\end{equation}
The upper bound~\eqref{kernel:intermidiary} of the kernel gives control over the next two terms:
\[
\left|H_d(\lambda k_1) k_1 f(k_1)\right| 
\leq c_d^+\pi^2  \lambda^2 k_1^3 f(k_1)
\]
and
\[
\left|H_d(\lambda k_2) k_2 f(k_2)\right| 
\leq c_d^+ k_2 f(k_2).
\]
Let us now compare those pointwise values of $f$ with $\mathbb{WK}_d[f]$.
Proposition~\ref{prop:powerlawconsequence} gives a lower bound of $f$ on the interval~$[k_1,k_2]$:
\[
f(k)\geq \left[\sup_{k_1\leq k_0\leq k_2}f(k_0)k_0^\alpha\right] k^{-\alpha} \exp\left(-\|f\|_{\power^{-\alpha}_{0}(k_1,k_2)}\right) \cdot \mathbf{1}_{[k_1,k_2]}(k).
\]
As the kernel of the Wiener-Kinchine transform is positive,
this provides a point-wise lower bound on $\mathbb{WK}_d[f]$:
\[
\mathbb{WK}_d[f](\lambda)\geq   
\left[\sup_{k_1\leq k_0\leq k_2}f(k_0)k_0^\alpha\right] 
 \exp\left(-\|f\|_{\power^{-\alpha}_{0}(k_1,k_2)}\right) \cdot 
\mathbb{WK}_d[k^{-\alpha} \mathbf{1}_{[k_1,k_2]}](\lambda).
\]
The last transform can be computed explicitly. Let us recall that $1<\alpha<3$. One has:
\begin{align*}
\mathbb{WK}_d[k^{-\alpha} \mathbf{1}_{[k_1,k_2]}](\lambda) &= \int_{k_1}^{k_2} H_d(\lambda k)k^{-\alpha} dk
= \lambda^{\alpha-1} \int_{\lambda k_1}^{\lambda k_2} H_d(\sigma) \sigma^{-\alpha} d\sigma\\
&\geq c_d^- \lambda^{\alpha-1}\int_{\lambda k_1}^{\lambda k_2} \frac{\pi^2 \sigma^{2-\alpha}}{1+\pi^2\sigma^2} d\sigma
\geq
\frac{1}{2}c_d^- \lambda^{\alpha-1} \left(
\pi^2 \int_{k_1\lambda}^{1/\pi} \sigma^{2-\alpha} d\sigma
+ \int_{1/\pi}^{k_2\lambda} \sigma^{-\alpha} d\sigma\right)\\
&\geq \frac{1}{2}c_d^- \lambda^{\alpha-1} \left(
\frac{2\pi^{\alpha-1}}{(3-\alpha)(\alpha-1)} 
- \frac{\pi^2}{3-\alpha} (k_1\lambda)^{3-\alpha} - \frac{1}{\alpha-1} (k_2\lambda)^{-(\alpha-1)}
\right).
\end{align*}
For $\lambda\in(\frac{\mu}{k_2},\frac{\varepsilon}{k_1})$ and using assumption~\eqref{epssmallmularge}, one gets:
\[
\mathbb{WK}_d[k^{-\alpha} \mathbf{1}_{[k_1,k_2]}](\lambda)  \geq 
c_d^-   \lambda^{\alpha-1} \left(
 \frac{\pi^{\alpha-1}}{(3-\alpha)(\alpha-1)}
 - \frac{\pi^2 \varepsilon^{3-\alpha}}{2(3-\alpha)}
 - \frac{\mu^{-(\alpha-1)}}{2(\alpha-1)}
 \right)
 \geq \frac{c_d^- \pi^{\alpha-1}}{2(3-\alpha)(\alpha-1)} \lambda^{\alpha-1}.
\]
Combining the previous inequality with either $k_0=k_1$ or $k_0=k_2$ provides upper bounds of~$f(k_i)$:
\[
c_d^+\pi^2\lambda^2 k_1^3 f(k_1) \leq 
2 \,\frac{c_d^+}{c_d^-}\, (3-\alpha)(\alpha-1) \exp\left(\|f\|_{\power^{-\alpha}_{0}(k_1,k_2)}\right)
\cdot (\pi \lambda k_1)^{3-\alpha}  \, \mathbb{WK}_d[f](\lambda),
\]
\[
c_d^+ k_2 f(k_2) \leq
2 \,\frac{c_d^+}{c_d^-}\, (3-\alpha)(\alpha-1) \exp\left(\|f\|_{\power^{-\alpha}_{0}(k_1,k_2)}\right)
\cdot (\pi \lambda k_2)^{-(\alpha-1)}  \, \mathbb{WK}_d[f](\lambda),
\]
and finally:
\begin{equation}\label{HRpointwisecontrol}
c_d^+\pi^2\lambda^2 k_1^3 f(k_1) + c_d^+ k_2 f(k_2) \leq 
4 \,\frac{c_d^+}{c_d^-}\,  \exp\left(\|f\|_{\power^{-\alpha}_{0}(k_1,k_2)}\right) \sigma_\alpha(\varepsilon,\mu) \, \mathbb{WK}_d[f](\lambda).
\end{equation}

\paragraph{Case $d\geq 3$.}
In this case, one can use $z |H'(z)|\leq 2 H(z)$ from Lemma~\ref{lemma:L}. Therefore:
\begin{align*}
\left|\int_0^{k_1} \left[\lambda k H_d'(\lambda k)-(\alpha-1)H_d(\lambda k)\right] f(k)dk\right|
&\leq (\alpha+1) \int_0^{k_1} H_d(\lambda k) f(k) dk\\
&\leq (\alpha+1) \, c_d^+\pi^2  \lambda^2 \int_0^{k_1}k^2 f(k) dk
\end{align*}
and similarly
\begin{align*}
\left|\int_{k_2}^\infty \left[\lambda k H_d'(\lambda k)-(\alpha-1)H_d(\lambda k)\right] f(k)dk\right|
&\leq (\alpha+1) \int_{k_2}^\infty H_d(\lambda k) f(k) dk\\
&\leq (\alpha+1) \, c_d^+ \int_{k_2}^\infty  f(k) dk.
\end{align*}
Assumption~\eqref{assumption:tail} allows us to convert the last remainder terms of~\eqref{HRdecomp}  into the ones we already dealt with:
\begin{equation}\label{HRccl}
\|\mathbb{WK}_d[f]\|_{\power^{\alpha-1}_\infty(\frac{\mu}{k_2},\frac{\varepsilon}{k_1})} 
\leq \|f\|_{\power^{-\alpha}_\infty(k_1,k_2)}
+ C \exp\left(\|f\|_{\power^{-\alpha}_{0}(k_1,k_2)}\right) \sigma_\alpha(\varepsilon,\mu)
\end{equation}
with 
\begin{equation}\label{HRconstDimd}
C = 4\left(1+(\alpha+1)(C_1\vee C_2)\right)\,\frac{c_d^+}{c_d^-}
\cdotp
\end{equation}

\paragraph{Case $d=2$.} 
Let us use the primitive $G_2(z)$ of $z H_2'(z)$ that we introduced in the proof of~\eqref{URinfCd2bis}.
One can easily check that $|G_2(z)| \leq z$ on $\R_+$. An integration by part followed by~\eqref{assumption:dim2} thus gives:
\[
\left|\int_{k_2}^\infty \lambda k H_d'(\lambda k) f(k) dk \right|
\leq k_2 f(k_2)
+ \int_{k_2}^\infty k |f'(k)| dk
\leq (1+C_2)k_2 f(k_2).
\]
For the other remainder, one uses that $z |H_2'(z)|\leq 4z^2$ on $\R_+$ followed
by~\eqref{assumption:tail}:
\[
\left|\int_0^{k_1} \lambda k H_d'(\lambda k) f(k) dk \right|
\leq 4 \lambda^2  \int_0^{k_1} k^2 f(k) dk
\leq 4 C_1 \lambda^2 k_1^3 f(k_1).
\]
Both terms are thus controlled by~\eqref{HRpointwisecontrol}. 
One thus gets~\eqref{HRccl} again but with the constant $C$ modified:
\begin{equation}\label{HRconstDim2}
C = 4\left(1+(\alpha-1)(C_1\vee C_2)
+ \frac{4C_1}{c_2^+ \pi^2} \vee \frac{1+C_2}{c_2^+}
\right)\,\frac{c_2^+}{c_2^-} \cdotp
\end{equation}

\paragraph{Case $d=1$.}
The proof is similar to that for $d=2$. One uses the fact that the function $G_1$ introduced in the proof of~\eqref{URinfCd1bis}
satisfies $G_1'(z)=z H_1'(z)$ and $|G_1(z)|\leq 2z$ on $\R_+$.  One also uses the fact that $z |H_1'(z)|\leq 4 \pi^2 z^2$ on~$\R_+$.
Thus~\eqref{HRccl} still holds with
\begin{equation}\label{HRconstDim1}
C = 4\left(1+(\alpha-1)(C_1\vee C_2)
+ \frac{4C_1}{c_1^+} \vee \frac{2(1+C_2)}{c_1^+}
\right)\,\frac{c_1^+}{c_1^-} \cdotp
\end{equation}
This concludes the proof of theorem~\ref{thm4}.
\cqfd

\section{Numerical examples (see \S\ref{par:numericsexplained})}
\label{par:numerics}

\newpage
\section{Appendix : properties of the Wiener-Khinchin kernel $H_d(\sigma)$}\label{par:appendix}

This section contains different asymptotic bounds of the Wiener-Khinchin kernel~\eqref{WKkernel}.
The constants $c_d^{\pm}$ and $c_d^0$, used throughout the article, are also defined here.
For further properties, see also \cite{MK72}.

\begin{figure}[H]
\captionsetup{width=.6\linewidth}
\begin{center}
\resizebox{.5\linewidth}{!}{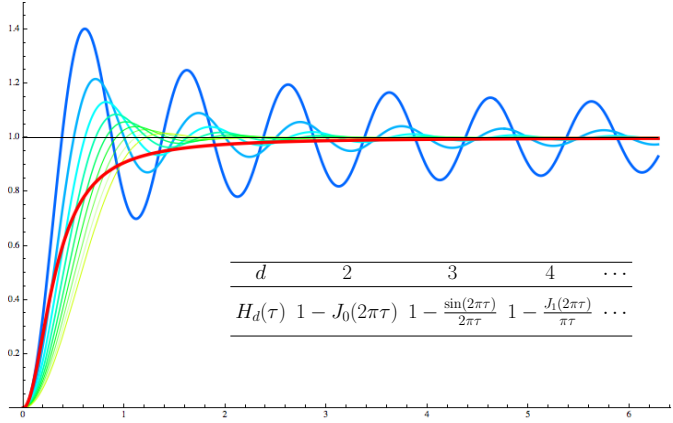}
\caption{\label{fig:kernel}\small
Kernels $H_d(\sigma)$ for $d=2,3,\ldots, 10$ on $[0,2\pi]$. The thickness of the drawing decreases with $d$ and the red line corresponds
to the approximation \eqref{kernel:intermidiary} \ie $\sigma^2/(\pi^{-2}+\sigma^2)$.}
\end{center}
\end{figure}

\medskip
For $d\geq 2$, the kernel of \eqref{main} is non-negative and converges to 1 in an oscillatory way. Figure~\ref{fig:kernel} shows the profile of $H_d$ for different values of $d$.
Dimension $d=1$ requires a special attention because the oscillations of $H_1(\sigma)=1-\cos(2\pi \sigma)$ are not damped;
however, $H_1(\sigma)\geq0$ still holds.

\begin{description}
\item[Rough bounds] For $d=1$, one has $H_1(\sigma)\leq c_1^+ = 2$.
For $d\geq 2$, one has:
\begin{equation}\label{kernel:intermidiary}
\forall \sigma\geq0,\qquad
c_d^-\cdot\frac{\pi^2\sigma^2}{1+\pi^2\sigma^2}\leq H_d(\sigma)\leq c_d^+ \cdot \frac{\pi^2\sigma^2}{1+\pi^2\sigma^2} \leq c_d^+
\end{equation}
with  $c_d^+\leq \frac{d}{d-1} \leq 2$ and $c_d^-\geq \begin{cases}3/4 & \text{if }d=2,\\ 2/d &  \text{if }d\geq 3.\end{cases}$

\begin{table}[H]
\captionsetup{width=.52\linewidth}
$$\begin{array}{cccccc}\hline
d & 2 & 3 & 4 & 5 & 6\\\hline
c_d^-  & 0.756 & 2/3 & 1/2 & 2/5 & 1/3 \\\hline
c_d^+ & 1.839  & 1.487    & 1.322   & 1.230 & 1.172\\\hline
c_d^+/c_d^- & 2.433  & 2.231    & 2.644   & 3.075 & 3.516\\\hline
\end{array}$$
\caption{\label{Table:cpm}\small
Numerical values for $c_d^-$ (rounded by default for $d=2$) and for $c_d^+$ and $c_d^+/c_d^-$ (rounded by excess).}
\end{table}

\item[Finer bound at the origin] For $\sigma\in [0,\sqrt{d+2}/\pi]$ and $d\geq 1$, one has:
\begin{equation}\label{kernel:origin}
H_d(\sigma) = \frac{2\pi^2}{d}\sigma^2 - \underline\varepsilon_d(\sigma) \qquad\text{with}\qquad
0\leq \underline\varepsilon_d(\sigma) \leq \frac{2\pi^4}{d(d+2)}\sigma^4.
\end{equation}
The estimate is actually valid for any $\sigma\geq0$ but the proposed upper bound of the remainder becomes useless for large $\sigma$.
\item[Finer bound at infinity] One can capture the oscillations at infinity in a very precise way. There exists a constant $c_d^0$ such that, for any $\sigma> 0$:
\begin{equation}\label{kernel:asymptotic}
H_d(\sigma) = 1 - \frac{\Gamma\left(\tfrac{d}{2}\right)}{\pi^{d/2}} \sigma^{-\frac{d-1}{2}} \cos\left[\left(2\sigma-\frac{d-1}{4}\right)\pi\right] + \bar\varepsilon_d(\sigma) \qquad\text{with}\qquad
| \bar\varepsilon_d(\sigma)| \leq c_d^0 \sigma^{-\frac{d+1}{2}}.
\end{equation}
For $d=1$ and $d=3$, this formula is actually exact, \ie $\bar\varepsilon_1(\sigma)=\bar\varepsilon_3(\sigma)\equiv0$. 
\begin{table}[H]
$$\begin{array}{cccccccccc}\hline
d & 1 & 2 & 3 & 4 & 5 & 6 & \ldots & 10 & 20 \\\hline
10^3 \times c_d^0 &  {\color{red}\mathbf{0}}  &  6.34 & {\color{red}\mathbf{0}} & 6.05 &  12.1 &  20.2 & \ldots & 101 & 24'900\\\hline
\end{array}$$
\caption{\small
Numerical values of $c_d^0$ (rounded by excess), as multiples of $10^{-3}$.}
\end{table}
One has the following estimate:
\begin{equation}\label{kernel:Hminus1}
|H_d(\sigma)-1|\leq (1+\mathbf{1}_{d\geq4})\Gamma\left(\tfrac{d}{2}\right)\pi^{-d/2} \sigma^{-\frac{d-1}{2}}.
\end{equation}
For $d=1,2,3$, \eqref{kernel:Hminus1} holds without restrictions on $\sigma\geq0$ ; for $d\geq4$, the estimate \eqref{kernel:Hminus1} holds for
$$\sigma\geq c_d^0 \cdot \pi^{d/2}/\Gamma\left(\tfrac{d}{2}\right).$$
Note that the absence of restrictions on $\sigma$ for $d=2$ follows from $\pi\sqrt{z}J_0(2\pi z)\leq 1$.
\end{description}

\bigskip
For the derivatives, one needs a control of the following quantity:
\begin{equation}
L_d(z)=2H_d(z)-z H_d'(z).
\end{equation}
The behavior of the kernel $L_d(z)$ is universally very good at the origin, but it degenerates at infinity when the dimension is small.

\begin{lemma}\label{lemma:L}
For any $d\geq1$, there exists a constant $C_L>0$ such that
\[
\forall z\in\R_+,\qquad |L_d(z)| \leq C_L \begin{cases}
\min\{ z^4 ; 1\} & \text{if }d\geq 3,\\
\min\{ z^4 ; z^{1/2}\} & \text{if }d=2,\\
\min\{ z^4 ; z\} & \text{if }d=1.
\end{cases}
\]
Moreover, when $d\geq3$, one has $L_d(z)\geq 0$ and even better:  $z |H_d'(z)|\leq 2H_d(z)$.
\end{lemma}

\begin{table}[H]
\captionsetup{width=.5\linewidth}
$$\begin{array}{ccccccc}\hline
d & 1 & 2 & 3 & 4 & 5 & 6  \\\hline
\multirow{2}{*}{$C_L=\frac{4\pi^4}{d(d+2)}$} & 4\pi^4/3  & \pi^4/2   & 4\pi^4/15 & \pi^4/6 & 4\pi^4/35   & \pi^4/12 \\\cline{2-7}
 & 129.9  & 48.8   & 26. & 16.3 & 11.2   & 8.2 \\\hline
\end{array}$$
\caption{\label{T:CL}\small
Values of $C_L$ (rounded by excess).}
\end{table}

\begin{lemma}\label{lemma:L2}
For any $\mu>0$, a more refined control of the derivatives is given by:
\[
\max_{z\in[0,\mu]} \left|\frac{z H_d'(z)}{H_d(z)}\right|=
\begin{cases}
2(1+\sqrt{\mu}) & \text{if } d=2,\\
2 & \text{if }d\geq3
\end{cases}
\]
and
\[
\max_{z\in[\mu,\infty]} \left|\frac{z H_d'(z)}{H_d(z)}\right|
=\begin{cases}
\infty & \text{if } d=2,\\
\leq 2 & \text{if }d=3,\\
\mathcal{O}(\mu^{-\frac{d-3}{2}}) & \text{if } d\geq4.
\end{cases}
\]
\end{lemma}

\paragraph{Proof of Lemma~\ref{lemma:L}}
Using the asymptotic expansion at the origin of the Bessel function provides:
\[
L_d(z)=\frac{4\pi^4}{d(d+2)} z^4 - \frac{16\pi^6}{3d(d+2)(d+4)} z^6 + \mathcal{O}(z^8),
\]
hence the behavior near the origin.
As $L_d$ is a smooth function, it is locally bounded on any compact set of $\R_+$.
For the behavior at infinity, one uses:
\[
L_d(z)= 2-\Gamma\left(\frac{d}{2}\right)(\pi z)^{1-\frac{d}{2}}\left[\left(\frac{d}{2}+1\right)J_{\frac{d}{2}-1}(2\pi z) +2\pi zJ_{\frac{d}{2}-1}'(2\pi z)\right]
\]
and the classical rough asymptotic expansion of the Bessel functions:
\[
J_n(t) \underset{t\to+\infty}{=} \mathcal{O}(t^{-1/2}).
\]
The derivatives satisfy the same asymptotic because $J_n'(t)=\frac{1}{2}\left(J_{n-1}(t)-J_{n+1}(t)\right)$ for $n\neq0$. In particular, for any $d\geq3$,
the function $L_d$ is bounded. For $d=3$, one has
\[
L_3(z)=2-\cos(2\pi z)-\frac{\sin(2\pi z)}{\pi z}
\]
and for $d\geq 4$, the function $L_d(z)$ converges to $2$ at infinity.
For the exceptions of dimension 1 and 2, it is simpler to compute the kernels explicitly:
\[
L_1(z) = 2-\cos(2\pi z)+2\pi z\sin(2\pi z) 
\]
\[
L_2(z)=2-2J_0(2\pi z)+2\pi z J_1(2\pi z).
\]
It is then quite clear that $L_1(z)= \mathcal{O}(z)$ and $L_2(z)= O (z^{1/2})$.
\cqfd

\paragraph{Proof of Lemma~\ref{lemma:L2}}
For the comparisons, one uses the following identities:
\[
\frac{z H_d'(z)}{H_d(z)} 
= \frac{4\pi^2 z^2}{d} \cdot \frac{ {}_0F_1\left(1+\frac{d}{2};-\pi^2 z^2\right) }{ 1- {}_0F_1\left(\frac{d}{2}; -\pi^2 z^2\right) }
= \frac{2 \Gamma\left(\frac{d}{2}\right) (\pi z)^2  J_{d/2}(2\pi z)}{(\pi z)^{d/2} - \Gamma\left(\frac{d}{2}\right)\pi z  J_{d/2-1}(2\pi z)}
\]
where ${}_0F_1$ is a confluent hypergeometric function:
\[
{}_0F_1(a,z) = \sum_{k=0}^\infty \frac{\Gamma(a)}{\Gamma(a+k)} \frac{z^k}{k!}\cdotp
\]
For $d\geq3$, the maximum value equals 2 and is achieved at the origin
and the decay at infinity follows from the asymptotic decay of the Bessel functions.
The case $d=2$ is an  exercice in calculus.
\cqfd


\bigskip
Finally, as the derivatives of the kernel behave worse in dimension 2 than in dimension 3, the following
comparison can be useful:
\begin{equation}\label{sandwich}
\forall z\geq0,\qquad \frac{3}{4} H_{3}(z) \leq H_2(z) \leq \frac{3}{2} H_{3}(z).
\end{equation}
More generally, one has:
\begin{equation}\label{sandwich2}
\forall z\geq0,\qquad \frac{d+1}{d+2} \, H_{d+1}(z) \leq H_d(z) \leq \frac{d+1}{d} \, H_{d+1}(z)
\end{equation}
for any dimension $d\geq2$.

\end{document}